\documentclass[11pt,oneside]{article}
\usepackage{a4wide}
\usepackage{mathrsfs}
\usepackage{latexsym,bm}
\usepackage{graphicx}
\usepackage{indentfirst}
\usepackage{slashed}
\usepackage{amsmath}
\usepackage{amssymb}
\usepackage{color}
\usepackage{hyperref}
\usepackage{epsfig}
\usepackage[titletoc]{appendix}
\usepackage{multirow}%
\usepackage{rotating}
\usepackage{cite}
\usepackage{ulem} 
\usepackage{extpfeil}
\usepackage[top=2.9cm,bottom=2.5cm,left=2.8cm,right=3cm]{geometry}
\usepackage{tabularx}

\newcommand{\email}[1]{\footnote{{\em } \texttt{#1}}}

\newcommand{\pim}{\pi^{-}}
\newcommand{\pin}{\pi^{0}}
\newcommand{\bra}{\langle}
\newcommand{\ket}{\rangle}
\newcommand{\mL}{\mathcal{L}}

\newcommand{\mM}{\mathcal{M}}

\newcommand{\newtex}[1]{{\color{black}{#1}}}

\begin{document}
\thispagestyle{empty}
\title{
\Large \bf Triple-product asymmetry in the radiative two-pion tau decay } 
\author{\small Cheng~Chen$^{a}$,\, Chun-Gui Duan$^{a}$,\,  Zhi-Hui Guo$^{a,b}$\email{zhguo@hebtu.edu.cn} \\[0.3em]
{ \small\it ${}^a$  Department of Physics and Hebei Advanced Thin Films Laboratory, } \\
{\small\it Hebei Normal University,  Shijiazhuang 050024, China}\\[0.1em]
{\small\it ${}^b$ School of Physics, Southeast University, Nanjing 211189, China } \\[0.1em]
}

\date{}

%

\maketitle
\begin{abstract}
In this work, we perform a detailed study of the $\tau^-\to\pi^-\pi^0\gamma\nu$ decay process within the resonance chiral theory. We pay special attention to the triple-product asymmetry in the $\tau^-\to\pi^-\pi^0\gamma\nu$ process. The minimal resonance chiral Lagrangian and the odd-intrinsic parity resonance operators are simultaneously included to calculate the decay amplitudes. Various invariant-mass distributions in the $\pi^-\pi^0$, $\pi^-\gamma$ and $\pi^0\gamma$ systems are studied and they reveal different resonance dynamics. We further predict the intriguing nonzero triple-product  asymmetry distributions, which may provide useful guidelines for future experimental measurements conducted at the Belle-II and super tau-charm facilities. 
\end{abstract}

\section{Introduction}

Charge-conjugation and parity violation (CPV) is one of the most important open problems in the Standard Model (SM) and beyond. The typical CPV observables are the decay rate asymmetries from the processes related with the charge conjugations, which have been widely used to establish the CPV in the strange-, beauty- and charm-meson sectors. Up to now the CPV in the lepton sector is yet to be discovered, and its establishment will definitely extend our understanding of the CPV in particle physics. The charge-conjugate decay rate asymmetries in various channels are expected to provide promising opportunities to establish the CPV of the $\tau$ lepton~\cite{Bigi:2005ts,BABAR:2011aa,Cirigliano:2017tqn,Datta:2006kd,Chen:2021udz,Chen:2019vbr}. Another distinct way to probe the CPV relies on the nontrivial kinematical measurements of the T-odd quantities. One of such observables is the triple-product asymmetry~\cite{Golowich:1988ig,Braguta:2001nz,Braguta:2003wf,Muller:2006gu,Gronau:2011cf,Rudenko:2011qe,Durieux:2015zwa}, which can be constructed via $\varepsilon_{\mu\nu\rho\sigma} p_1^\mu p_2^\nu p_3^\rho p_4^\sigma$, being $\varepsilon_{\mu\nu\rho\sigma}$ Levi-Civita anti-symmetric tensor and $p_{i=1,2,3,4}$ the momenta of the involved particles. Other types of triple-product asymmetry quantities with spin vectors, which usually requires to measure the polarizations of the final-state particles, can be also constructed in a similar way~\cite{Bigi:2011em,Datta:2006kd}.

The T-odd kinematical variable $\xi\equiv\varepsilon_{\mu\nu\rho\sigma} p_1^\mu p_2^\nu p_3^\rho p_4^\sigma$ naturally originates from the four-body decay processes. In the rest frame of the decaying particle, without loss of any generality, the quantity $\xi$ can be written as a term that is proportional to $(\vec{p}_1\times \vec{p}_2)\cdot \vec{p}_3$, which gives rise to the name of the triple product for such observables. The triple-product asymmetries have been extensively investigated in the charged and neutral $K_{l\,3\gamma}$ decays, $i.e.$, $K^{+}\to\pin l^+ \nu_l \gamma$ and $K^{0}\to\pim l^+ \nu_l \gamma$~\cite{Braguta:2001nz,Braguta:2003wf,Muller:2006gu,Rudenko:2011qe}. It is found that the electromagnetic final-state interactions from the photon loops in the SM~\cite{Braguta:2001nz,Rudenko:2011qe}, rather than the strong interactions from the hadronic loops~\cite{Muller:2006gu}, give the dominant contributions to the triple-product asymmetries in the $K_{l\,3\gamma}$. There are two main effects that reduce the hadronic contributions in the $K_{l\,3\gamma}$ decay processes. First, the nonvanishing hadronic effects in the triple-product asymmetries only enter in the structure dependent (SD) parts of the $K_{l\,3\gamma}$ decay matrix elements, which are much suppressed than the inner bremsstrahlung or structure independent (SI) parts. Second, the hadronic contributions starting from the two-pion threshold are reduced to a great extent due to the small kinematical phase space up to the kaon mass. However, these two suppression conditions do not hold in the $\tau\to\pi\pi\gamma\nu_{\tau}$ decay. To be more specific, not only the hadronic effects in the SD part but also their effects in the SI part can contribute to the triple-product asymmetry distributions in the radiative two-pion $\tau$ decays. In addition, there is no phase space suppression for the hadronic contributions in the $\tau\to\pi\pi\gamma\nu_{\tau}$ process. Therefore, the hadronic contributions are expected to play significant roles in the triple-product asymmetries in the aforementioned two-pion radiative $\tau$ decay, a feature that is rather different from the $K_{l\,3\gamma}$ decays. 

It should be noted that both the CP-conserving (CPC) and CPV interactions can cause the triple-product asymmetries. The genuine CPV interactions can be probed by taking the differences of the triple-product asymmetries resulting from the charge-conjugate processes. In another word, for the $\tau\to\pi\pi\gamma\nu_{\tau}$ processes, the triple-product asymmetry arising from one charged state of $\tau$ is contributed by both the SM dominant CPC interactions and the possible beyond SM (BSM) dominant CPV ones. 
As an exploratory study, we focus on the triple-product asymmetry in the $\tau^{-}\to\pi^{-}\pi^{0}\gamma\nu$ process that originates from the final-state strong interactions in the SM. It turns out that such asymmetry can provide a useful object to discriminate different hadronic models and inputs. Our work also paves the way to a future study of the genuine CPV signals by  studying triple-product asymmetries in the two-pion radiative $\tau$ decay. 
According to the CPV study in the $K_{l3\gamma}$~\cite{Braguta:2003wf,Muller:2006gu}, the genuine CPV signals from the charge-conjugation $\tau\to\pi\pi\gamma\nu$ processes will be also proportional to the various combinations of hadronic form factors in the CPC sector. Although the sizable hadronic uncertainties may affect the precise determination of the CPV distributions in $\tau\to\pi\pi\gamma\nu$, genernally speaking, the large hadronic contributions are expected to enhance the CPV signals in the radiative two-pion tau decays.

This paper is organized as follows. In Sec.~\ref{sec.theo}, we introduce the relevant resonance chiral Lagrangians and perform the calculations of the decay amplitudes of the $\tau^-\to\pi^-\pi^0\gamma\nu_\tau$ process. The general  discussions on the triple-product asymmetries in the radiative two-pion tau decays are given in detail in Sec.~\ref{sec.todd}. The phenomenological studies, including the determinations of the free couplings, the sensitivity of the $\tau\to\pi\pi\gamma\nu_\tau$ branching ratios on the photon energy cutoffs, the discussions of the resonance dynamics in various two-particle invariant mass distributions, are carried out in Sec.~\ref{sec.pheno}, in which we also give the promising  predictions of the nonzero triple-product asymmetry distributions. A short summary and conclusions are then presented in Sec.~\ref{sec.conclusion}. Essential technical details, including the treatment of phase spaces, and the explicit formulas of the relevant form factors and the kinematical coefficients when evaluating the triple-product asymmetries are relegated to the appendices.

\section{Resonance chiral Lagrangians and relevant amplitudes of the radiative two-pion tau decay}\label{sec.theo}
 
The matrix element of the $\tau^-(P)\to\pim(p_1)\pin(p_2)\nu_\tau(q)\gamma(k)$ decay process can be generally written as~\cite{Cirigliano:2002pv} 
\begin{eqnarray}\label{eq.defT}
\mM= e\,G_F\,V_{ud}^* \epsilon^{*\mu}(k) \!\!\!\!\!\!\!\!\!\! && \bigg\{  F_\nu \bar{u}(q)\gamma^\nu(1-\gamma_5)(m_\tau+\slashed{P}-\slashed{k})\gamma_\mu u(P) \nonumber  \\
&&+(V_{\mu\nu}-A_{\mu\nu})\bar{u}(q)\gamma^\nu(1-\gamma_5)u(P) \bigg\}\,, \label{eq.fullamp}
\end{eqnarray}
where $e$ is basic unit of electric charge, $G_F$ stands for the Fermi constant, $V_{ud}$ denotes the CKM matrix element and $\epsilon^{*\mu}$ corresponds to the polarization vector of the photon field. 
\newtex{The effects from the bremsstrahlung off the $\tau$ lepton are encoded in the $F_\nu$ term, which is governed by the non-radiative decay process, as illustrated in Fig.~\ref{fig.fv}. Its explicit expression is given by 
\begin{equation}\label{eq.Fnu}
F_\nu=(p_2 -p_1)_\nu F_V(t)/(2P\cdot k) \,,
\end{equation}
where $t=(p_1+p_2)^2$ and $F_V(t)$ stands for the vector form factor of the two pions~\cite{Guerrero:1997ku,GomezDumm:2013sib}. In the isospin limit, the charged form factor $F_V(t)$ can be directly related to the electromagnetic form factor of the pions.

\begin{figure}[htbp]
\begin{center}
\includegraphics[angle=0, width=0.5\textwidth]{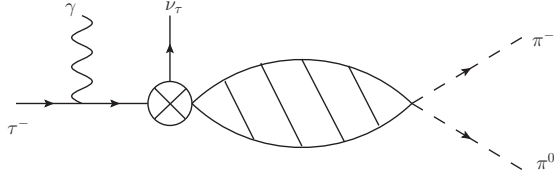}
\caption{ Relevant Feynman diagram to the bremsstrahlung off the initial $\tau$ lepton. It is completely determined by the pion vector form factor $F_V(t)$ that governs the non-radiative $\tau\to\pi\pi\nu_\tau$ decay process. The vector form factor $F_V(t)$ includes the contributions from the nonperturbative two-pion strong interactions~\cite{Guerrero:1997ku,GomezDumm:2013sib}, denoted by the slashed bubble. \label{fig.fv}}
\end{center}
\end{figure}

The remaining hadronic tensor amplitudes $V_{\mu\nu}$ and $A_{\mu\nu}$ in Eq.~\eqref{eq.fullamp} encode the dynamics of the transition $W^-\to \pi^-\pi^0\gamma$. Both the contributions from the bremsstrahlung off the $\pim$ and the SD parts enter in the $V_{\mu\nu}$ and $A_{\mu\nu}$ terms. The requirement of gauge invariance on the matrix elements implies that
\begin{eqnarray} \label{eq.gaugev}
k^\mu V_{\mu\nu}  &=&  (p_2 -p_1)_\nu \,F_V(t)\,, \\
k^\mu  A_{\mu\nu} &=& 0 \,, \label{eq.gaugea}
\end{eqnarray}
which give practical constraints to write the explicit decompositions of such tensor amplitudes.  
For the vector amplitude $V_{\mu\nu}$, we take the same form as proposed in Ref.~\cite{Cirigliano:2002pv} 
\begin{eqnarray}\label{eq.Vmunu}
V_{\mu\nu}=\!\!\!\!\!\!\!\!\!\! && F_V(u)\frac{p_{1\mu}}{p_1 \cdot k}(p_1 +k-p_2)_\nu -F_V(u)g_{\mu\nu}  +\frac{F_V(u)-F_V(t)}{(p_1 +p_2)\cdot k}(p_1 +p_2)_\mu (p_2 -p_1)_\nu \nonumber \\
&&+v_1(g_{\mu\nu}p_1 \cdot k-p_{1\mu}k_\nu)+v_2(g_{\mu\nu}p_2 \cdot k-p_{2\mu}k_\nu) +v_3(p_{1\mu}p_2 \cdot k-p_{2\mu}p_1 \cdot k)p_{1\nu} \nonumber \\
&&+v_4(p_{1\mu}p_2 \cdot k-p_{2\mu}p_1 \cdot k)(p_1 +p_2 +k)_\nu \,,
\end{eqnarray}
with $u=(P-q)^2$. Under this decomposition form, every $v_{i=1\cdots 4}$ term fulfills the constraint in Eq.~\eqref{eq.gaugea}. 
It is easy to demonstrate~\cite{Cirigliano:2002pv} that the decomposition of $V_{\mu\nu}$ in Eq.~\eqref{eq.Vmunu} manifestly satisfies the Low's theorem in the soft-photon limit~\cite{Low:1958sn}. For a radiative decay process, the bremsstrahlung off the charged particles will lead to infrared divergences when the energy of the photon $E_\gamma$ approaches to zero. In the present tensor decomposition, the contributions from the bremsstrahlung off the $\tau$ and $\pi^-$ are given by the $F_\nu$ term~\eqref{eq.Fnu} and the first line in $V_{\mu\nu}$~\eqref{eq.Vmunu}, respectively, which will cause infrared divergences in the soft-photon limit. This also implies that the decay rate of the $\tau\to\pi\pi\gamma\nu_\tau$ is overwhelmingly dominated by the very low energy photons. In order to probe the interesting nontrivial dynamics in the radiative decay processes, one usually needs to introduce the photon energy cuts.  For the remaining $v_{i=1\cdots 4}$ terms in $V_{\mu\nu}$~\eqref{eq.Vmunu} and the $A_{\mu\nu}$ amplitude~\eqref{eq.Amunu}, they are free of the infrared singularity. 

For the axial-vector amplitude $A_{\mu\nu}$, there are four  independent form factors~\cite{Bijnens:1992en} and for later convenience in our calculation we parameterize them as 
\begin{equation}\label{eq.Amunu}
A_{\mu\nu}=i(a_1 \epsilon_{\mu\nu\rho\sigma}p_1^\rho k^\sigma+a_2 \epsilon_{\mu\nu\rho\sigma}p_2^\rho k^\sigma +a_3 p_{1\nu} \epsilon_{\mu\rho\beta\sigma}k^\rho p_1^\beta p_2^\sigma+a_4 p_{2\nu} \epsilon_{\mu\rho\beta\sigma}k^\rho p_1^\beta p_2^\sigma) \,,
\end{equation}
where the Schouten's identify has been used to write this decomposition. It is easy to check that each of the $a_{i=1\cdots 4}$ term fulfills the constraint in Eq.~\eqref{eq.gaugea}. It is noted that  somewhat different decompositions of the axial-vector amplitudes are used in Refs.~\cite{Bijnens:1992en,Miranda:2020wdg,Guevara:2016trs}. 
The pertinent Feynman diagrams arising in the minimal resonance chiral Lagrangian have been calculated in Ref.~\cite{Cirigliano:2002pv}, see Fig.~1 of the former reference for the $A_{\mu\nu}$ and Fig.~6 for the $V_{\mu\nu}$. We will pursue the calculations in resonance chiral theory by including the odd-intrinsic parity operators in this work, and the relevant Feynman diagrams originated from these operators to the $V_{\mu\nu}$ and $A_{\mu\nu}$ amplitudes will be illustrated in the following discussions. }

In the SM, the various transition form factors in Eqs.~\eqref{eq.Fnu}, \eqref{eq.Vmunu} and \eqref{eq.Amunu} are mainly governed by the nonperturbative strong interactions. Although chiral  perturbation theory can be used to reliably calculate the form factors in the very low energy region around the thresholds, it becomes inadequate in the energy region where hadron resonances start to contribute. Alternatively, resonance chiral theory (R$\chi$T)~\cite{Ecker:1988te,Ecker:1989yg}, which explicitly includes the bare resonance fields in a chiral covariant way, offers a reliable theoretical framework to calculate the form factors both in the low and resonant energy regions. Indeed R$\chi$T has been widely employed to investigate many phenomenological processes involving the light-flavor hadron resonances, including the meson-meson scattering~\cite{Bernard:1991zc,Oller:1998zr,Guo:2007ff,Guo:2011pa,Guo:2012ym,Guo:2012yt}, the hadronic $\tau$ decays~\cite{Dumm:2009va,Guo:2010dv,Nugent:2013hxa,Escribano:2014joa,Guevara:2016trs}, the form factors~\cite{Rosell:2004mn,Pich:2008jm}, the hadronic and radiative decay processes of various hadrons~\cite{Escribano:2010wt,Chen:2012vw,Chen:2013nna,Chen:2014yta}, etc. We further proceed the applications of the R$\chi$T to calculate the form factors appearing in the $\tau\to\pi\pi\gamma\nu_\tau$ process.

The minimal interaction operators with the vector and axial-vector resonances in R$\chi$T read~\cite{Ecker:1988te}
\begin{eqnarray}\label{eq.lagv2}
\mL_{V} &=&\frac{F_V}{2\sqrt{2}} \bra \hat{V}_{\mu\nu} f_{+}^{\mu\nu} \ket + i\frac{G_V}{\sqrt{2}} \bra \hat{V}_{\mu\nu} u^{\mu} u^{\nu} \ket\,, \\
\mL_{A} &=&\frac{F_A}{2\sqrt{2}} \bra \hat{A}_{\mu\nu} f_{-}^{\mu\nu} \ket \,.
\label{eq.laga2}
\end{eqnarray}
We refer to Ref.~\cite{Ecker:1988te} for further details about the definitions of the basic chiral building tensors $f_{\pm}^{\mu\nu}, u^\mu$ and the assignments of the flavor contents of the resonance multiplets $\hat{V}(\hat{A})_{\mu\nu}$. The contributions from the minimal R$\chi$T to the $\tau\to\pi\pi\gamma\nu_\tau$ decay amplitudes have been evaluated in Ref.~\cite{Cirigliano:2002pv}. Later on, a vector-meson-dominant (VMD) model was developed to include the refined effects from the odd-parity-intrinsic interacting vertex with $\rho\omega\pi$, together with other terms beyond the minimal R$\chi$T operators~\cite{FloresTlalpa:2005fz,FloresBaez:2006gf}. Compared to the results in Ref.~\cite{Cirigliano:2002pv}, the deviations obtained in  the latter two references are overwhelmingly caused by the intermediate $\omega$ meson. This conclusion is also confirmed by a recent study in Ref.~\cite{Miranda:2020wdg}, which extends the minimal R$\chi$T study by including a large amount of higher order odd- and even-intrinsic parity operators from Refs.~\cite{Kampf:2011ty,Cirigliano:2006hb}, respectively.

Although the high-energy behavior constraints help to substantially reduce the unknown couplings associated with the interaction operators, many of them are still undetermined and the authors of Ref.~\cite{Miranda:2020wdg} rely on the chiral counting arguments to roughly estimate the order of their magnitudes to proceed with the phenomenological studies. The loosely estimation of the free resonance couplings somewhat hinders the precise phenomenological discussions of the $\tau\to\pi\pi\gamma\nu_\tau$ decay~\cite{Miranda:2020wdg}. 
In the current work, we try to take a different strategy to present more definite phenomenological discussions. For the odd-intrinsic parity part, we will include the operators with both the $VVP$ and $VJP$ types within the framework of R$\chi$T~\cite{RuizFemenia:2003hm}, being $V$ the vector resonances, $P$ the light pseudo-scalar mesons and $J$ the external sources. In such a way, the additional contributions from the hadronic interaction vertices of the $\rho\pi\gamma, \omega\pi\gamma$ and $\omega\rho\pi$ types can be included, compared to the minimal R$\chi$T calculation in Ref.~\cite{Cirigliano:2002pv}. For the even parity sector, we will stick to the operators in Eqs.~\eqref{eq.lagv2} and \eqref{eq.laga2}  and refrain from including other types of interaction vertices beyond the minimal R$\chi$T, such as the $a_1\rho\pi$ ones, which usually introduce many free parameters. Due to the rather broad width of the $a_1$, one does not expect obvious resonance peaks in the $\tau\to\pi\pi\gamma\nu_\tau$ decays corresponding to the $a_1$ state. To account for the uncertainties of the lack of interaction vertices involving $a_1$ beyond the minimal R$\chi$T, the value of the coupling $F_A$ in  Eq.~\eqref{eq.laga2} that describes the interaction of the $a_1\pi\gamma$ will be varied, as shown in details in the phenomenological discussions later. This can be considered as a compensation to estimate the effects of omitting the higher-order hadronic interactions in the even parity sector. 

In this work, different from the resonance operator basis used in Ref.~\cite{Miranda:2020wdg}, we employ the ones proposed in Ref.~\cite{RuizFemenia:2003hm} to include the relevant odd-intrinsic parity operators beyond the minimal ones in the R$\chi$T framework. The merit is that the resonance Lagrangian in the latter basis has been widely exploited to investigate various physical processes and the relevant resonance couplings are also well determined~\cite{Chen:2012vw,Chen:2013nna,Chen:2014yta}, which can provide us valuable inputs for the study of the $\tau\to\pi\pi\gamma\nu_\tau$ decay. To further take the on-shell approximations of the $VJP$ operators and use the $\omega\to\pi^0\pi^0\gamma$ decay width as an additional input, we are able to make parameter-free predictions for the $\tau\to\pi\pi\gamma\nu_\tau$ decay. Therefore our study offers complementary results to the phenomenological aspects of the $\tau\to\pi\pi\gamma\nu_\tau$ process, compared to the ones given in Refs.~\cite{Cirigliano:2002pv,FloresTlalpa:2005fz,Miranda:2020wdg}. Furthermore, another important novelty of our work includes the exploratory discussions of the triple-product asymmetry distributions in the $\tau\to\pi\pi\gamma\nu_\tau$ process, which can supply an important guide for future experiment measurements.

The odd-intrinsic-parity Lagrangians consist of the $VVP$ and $VJP$ types. Their explicit forms are given by~\cite{RuizFemenia:2003hm}
\begin{eqnarray}\label{eq.lagvvp}
\mL_{VVP}=&& d_1 \varepsilon_{\mu\nu\rho\sigma} \langle \{V^{\mu\nu}, V^{\rho\alpha}\} \nabla_{\alpha}u^{\sigma}  \rangle
+ i d_2 \varepsilon_{\mu\nu\rho\sigma} \langle \{V^{\mu\nu}, V^{\rho\sigma}\} \chi_- \rangle
\nonumber \\ && + d_3 \varepsilon_{\mu\nu\rho\sigma} \langle \{ \nabla_\alpha V^{\mu\nu}, V^{\rho\alpha}\} u^{\sigma}  \rangle
+ d_4 \varepsilon_{\mu\nu\rho\sigma} \langle \{ \nabla^\sigma V^{\mu\nu}, V^{\rho\alpha}\} u_{\alpha}  \rangle \,,
\end{eqnarray}
and
\begin{eqnarray}\label{eq.lagvjp}
\mL_{VJP}=&& \frac{c_1}{M_V} \varepsilon_{\mu\nu\rho\sigma} \langle \{ V^{\mu\nu}, f_+^{\rho\alpha} \} \nabla_{\alpha}u^{\sigma}  \rangle
+\frac{c_2}{M_V} \varepsilon_{\mu\nu\rho\sigma} \langle \{ V^{\mu\alpha}, f_+^{\rho\sigma} \} \nabla_{\alpha}u^{\nu}  \rangle
 \nonumber \\ && + \frac{i c_3}{M_V} \varepsilon_{\mu\nu\rho\sigma} \langle \{ V^{\mu\nu}, f_+^{\rho\sigma}\} \chi_- \rangle
+\frac{i c_4}{M_V} \varepsilon_{\mu\nu\rho\sigma} \langle V^{\mu\nu} [ f_-^{\rho\sigma}, \chi_+ ] \rangle
\nonumber \\ &&+ \frac{c_5}{M_V} \varepsilon_{\mu\nu\rho\sigma} \langle \{ \nabla_{\alpha} V^{\mu\nu}, f_+^{\rho\alpha}\} u^{\sigma}  \rangle
+ \frac{c_6}{M_V} \varepsilon_{\mu\nu\rho\sigma} \langle \{ \nabla_{\alpha} V^{\mu\alpha}, f_+^{\rho\sigma}\} u^{\nu}  \rangle
\nonumber \\ &&+ \frac{c_7}{M_V} \varepsilon_{\mu\nu\rho\sigma} \langle \{ \nabla^{\sigma} V^{\mu\nu}, f_+^{\rho\alpha}\} u_{\alpha}  \rangle \,.
\end{eqnarray}

\begin{figure}[htbp]
\begin{center}
\includegraphics[angle=0, width=0.8\textwidth]{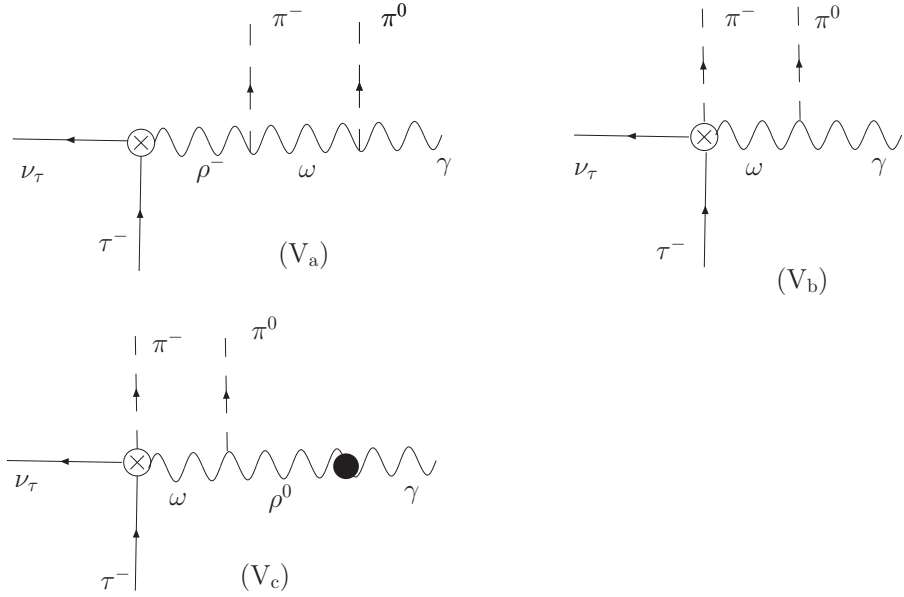}
\caption{ Feynman diagrams contributing to the vector form factors from the resonance chiral Lagrangians in Eqs.~\eqref{eq.lagvvp} and \eqref{eq.lagvjp}. \label{fig.v}}
\end{center}
\end{figure}

The Feynman diagrams that originate from the minimal R$\chi$T Lagrangians in Eqs.~\eqref{eq.lagv2} and \eqref{eq.laga2} are given in the Appendix of Ref.~\cite{Cirigliano:2002pv}. 
The relevant Feynman diagrams contributing to the vector and axial-vector form factors can be seen in Figs.~\ref{fig.v} and \ref{fig.a}, respectively. We give the explicit expressions of the vector and axial-vector form factors in the Appendix~\ref{sec.ff}. 
\newtex{As can be seen in those form factors, only some specific combinations of these resonance couplings appear in the final amplitudes. It turns out that many of such combinations can be fixed through the high-energy constraints that will be addressed in the following discussions.}

\begin{figure}[htbp]
\begin{center}
\includegraphics[angle=0, width=0.8\textwidth]{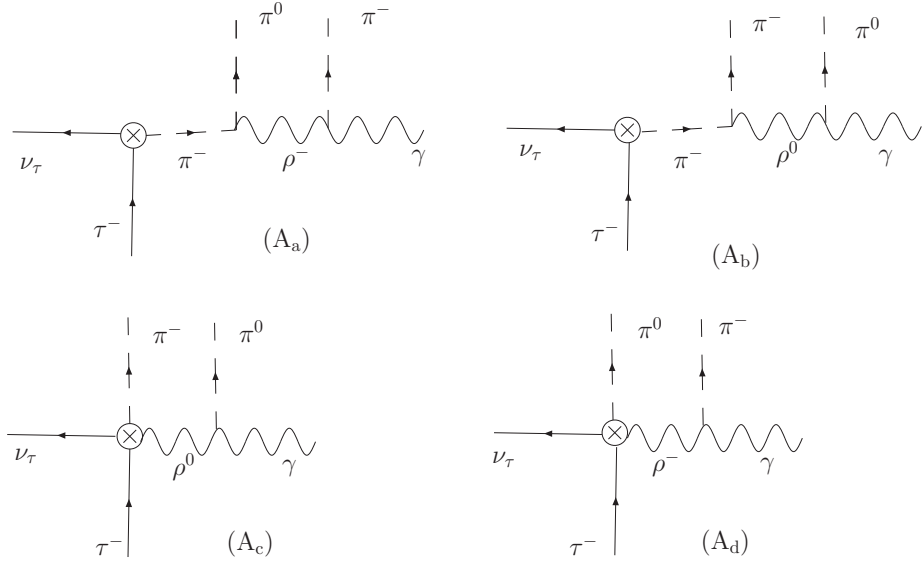}
\caption{Feynman diagrams contributing to the axial-vector form factors from the resonance chiral Lagrangians in Eqs.~\eqref{eq.lagvvp} and \eqref{eq.lagvjp}. \label{fig.a}}
\end{center}
\end{figure}

\section{Triple-product asymmetries in tau decays}\label{sec.todd}

The triple-product asymmetries can be constructed from the differential decay widths of the $\tau\to\pi\pi\gamma\nu_{\tau}$. After the sum/average of the spins of the particles in the final/initial states in Eq.~\eqref{eq.defT}, one can obtain the matrix element squared 
\begin{eqnarray}\label{eq.m2}
\frac{1}{2} \sum_{spins} |\mM|^2= \hat{M}_0 + \xi \hat{M}_1\,,
\end{eqnarray}
where both $\hat{M}_0$ and $\hat{M}_1$ are functions of the scalar products of the momenta of final-state particles, and the T-odd kinematical variable $\xi$ takes the form  
\begin{equation}\label{eq.defxi}
 \xi=\varepsilon_{\mu\nu\rho\sigma} P^\mu k^\nu p_1^\rho p_2^\sigma /m_\tau^4 \xlongequal[{\rm of}\,\tau]{{\rm rest\,frame }}  \vec{k} \cdot (\vec{p}_1 \times \vec{p}_2)/m_\tau^3 \,.
\end{equation}
It is noted that any even power of $\xi$ can be expressed in terms of the Lorentz scalar products of the final-state momenta, which further implies that $\hat{M}_0$ and $\hat{M}_1$ only depend on $\xi$ with even powers.

The differential decay width of the $\tau^-\to\pi^-\pi^0\gamma\nu_{\tau}$ process with respect to $\xi$ can be obtained by integrating out other kinematical variables of the invariant amplitude squared in Eq.~\eqref{eq.m2}. The triple-product asymmetry can be defined as 
\begin{equation}\label{asymmetry}
A_\xi = \frac{\Gamma_+ - \Gamma_-}{\Gamma_+ + \Gamma_-}\,,
\end{equation}
with 
\begin{equation}\label{eq.gammapm}
 \Gamma_{+} = \frac{(2\pi)^4}{2m_\tau} \int_{\xi>0} {\rm d}\Phi\, (\hat{M}_0+\xi\hat{M}_1 ) \,,\qquad 
 \Gamma_{-} = \frac{(2\pi)^4}{2m_\tau} \int_{\xi<0} {\rm d}\Phi\, (\hat{M}_0+\xi\hat{M}_1 ) \,.
\end{equation}
Clearly, the asymmetry $A_\xi$ is nonzero only when the function $\hat{M}_1$ is nonvanishing. Both the CP conserving final-state strong interactions and the CPV sector can contribute to $\hat{M}_1$. One can parameterize the two parts as 
\begin{equation}\label{eq.mhat1}
 \hat{M}_1 = \hat{M}_1^{\rm CP-EVEN} + \hat{M}_1^{\rm CP-ODD} \,,
\end{equation}
which transform under CP conjugation as 
\begin{equation}\label{eq.mhat1bar}
 \hat{M}_1 \to \overline{\hat{M}_1} = \hat{M}_1^{\rm CP-EVEN} - \hat{M}_1^{\rm CP-ODD}\,.
\end{equation}
The dominant effects of $\hat{M}_1^{\rm CP-EVEN}$ are from the SM and the most significant contributions to $\hat{M}_1^{\rm CP-ODD}$ are most  likely from the beyond SM sector. 
One possible way to factor out the different origins of the nonvanishing $\hat{M}_1$ is to simultaneously analyze $A_\xi$ and its counter part from the charge-conjugate process $\tau^+\to\pi^+\pi^0\gamma\nu_{\tau}$. We introduce a bar on top of each CP-conjugate quantity throughout. For example, $\overline{\xi}$, obtained by taking the CP transformation on $\xi$, changes it sign, due to the fact
\begin{equation}\label{eq.xibar}
\overline{\xi}= (-\vec{k}) \cdot [(-\vec{p}_1) \times (-\vec{p}_2)]/m_\tau^3 =-\xi\,. 
\end{equation}
Similarly one could define the triple-product asymmetry in the charge-conjugate process as 
\begin{equation}
\overline{A}_{\overline{\xi}} = \frac{\overline{\Gamma}_+ - \overline{\Gamma}_-}{\overline{\Gamma}_+ + \overline{\Gamma}_-}\,,
\end{equation}
with 
\begin{equation}\label{eq.gammapmbar}
 \overline{\Gamma}_{+} = \frac{(2\pi)^4}{2m_\tau} \int_{\overline{\xi}>0} {\rm d}\Phi\, (\overline{\hat{M}}_0+\overline{\xi}\overline{\hat{M}}_1 )\,,\qquad 
 \overline{\Gamma}_{-} = \frac{(2\pi)^4}{2m_\tau} \int_{\overline{\xi}<0} {\rm d}\Phi\, (\overline{\hat{M}}_0+\overline{\xi}\overline{\hat{M}}_1 )\,.
\end{equation}

In this work we focus on the asymmetries originated from the kinematical variable of the $\xi$ term, the CPV effects in the $\hat{M}_0$ part can be safely neglected, $i.e.$, we will take $\hat{M}_0=\overline{\hat{M}}_0$ throughout. In this case, it is easy to demonstrate that $\Gamma_++\Gamma_- =\overline{\Gamma}_+ + \overline{\Gamma}_-$. As mentioned previously, the asymmetries of $A_\xi$ and  $\overline{A}_{\overline{\xi}}$ receive contributions from both the CP-conserving and CPV effects. 
Then the difference and sum of the asymmetries from the charge-conjugate processes, $i.e.$, 
\begin{eqnarray}\label{eq.aaxi}
\mathcal{A}_\xi= A_\xi - \overline{A}_{\bar\xi}\,, 
\end{eqnarray}
and 
\begin{eqnarray}
 \mathcal{S}_\xi= A_\xi + \overline{A}_{\bar\xi}\,,
\end{eqnarray}
can be used to discern the CPV and CP-conserving dynamics in the $\tau\to\pi\pi\gamma\nu_\tau$ decays, respectively. 
This further indicates that if just one charged state of the $\tau$ is measured, instead of the two processes related with charge conjugations, one needs to subtract the CP-conserving contributions from the triple-product asymmetry $A_\xi$, in order to determine the CPV strength. As an exploring study, we focus on the triple-product asymmetry caused by the CP-conserving final-state strong interactions in this work. We leave the study of the CPV effects in the $\xi$ distributions in the $\tau\to\pi\pi\gamma\nu_\tau$ for a future work. 

\newtex{
The explicit expressions of $\hat{M}_0$ and $\hat{M}_1$ can be written in terms of the form factors in Eqs.~\eqref{eq.Fnu}, \eqref{eq.Vmunu} and \eqref{eq.Amunu} 
\begin{eqnarray}\label{eq.m02}
N^{-1}\hat{M}_0= \sum_{f,f'} C_{f f'}\,\mathrm{Re} (f^* f') + \sum_{f,I} C_{f,I}\, \mathrm{Re} (f^* I) + \sum_{I,I'} C_{I I'}\mathrm{Re} (I^* I')  \,,
\end{eqnarray}
\begin{eqnarray}\label{eq.m12}
N^{-1}\hat{M}_1=\sum_{f,f'} \tilde{C}_{f f'}\,\mathrm{Im} (f^* f') + \sum_{f,I} \tilde{C}_{f,I}\, \mathrm{Im} (f^* I) + \sum_{I,I'} \tilde{C}_{I I'}\mathrm{Im} (I^* I')  \,,
\end{eqnarray} 
where the normalization factor is $N=16m_\tau^2 e^2 G_F^2 V_{ud}^{2}$,  $f$ or $f'$ correspond to the pion vector form factors as both functions of $t$ and $u$, i.e.  $F_V(t)$ and $F_V(u)$, $I$ or $I'$ denote the SD vector and axial-vector form factors $v_{i=1,2,3,4}$ and $a_{i=1,2,3,4}$. The remaining terms $C_{f f'}$, $C_{f I}$, $C_{I I'}$, $\tilde{C}_{f f'}$, $\tilde{C}_{f I}$, $\tilde{C}_{I I'}$ are the kinematical factors, which are rather lengthy and explicitly given in the Appendix~\ref{sec.kinf}. }
In the $K\to\pi l\nu_l \gamma$ decay, the bremsstrahlung terms are governed by the $K\to\pi l \nu_l$ form factors $f_+$ and $f_1$, both of which are real in the entire physical energy ranges~\cite{Muller:2006gu}. As a result, the only nonvanishing strong interactions contribute to the triple-product asymmetry $A_\xi$ through the SD form factors $v_i$ and $a_i$, which turn out to be much suppressed~\cite{Muller:2006gu}, compared to the electromagnetic final-state interactions from the photon loops~\cite{Braguta:2001nz,Rudenko:2011qe}. In contrast, the pion vector form factors entering in the bremsstrahlung terms from the $\tau\to\pi\pi\nu_\tau\gamma$ process do contain significant nonvanishing imaginary parts. As a result, the final-state strong interactions are expected to provide important contributions to the triple-product asymmetry $A_\xi$ in the radiative two-pion tau decay process, a feature that is rather different from the $K_{l3\gamma}$ decays~\cite{Muller:2006gu,Braguta:2001nz,Rudenko:2011qe}.

\section{Phenomenological discussions}\label{sec.pheno}

In order to proceed with the phenomenological discussions, we need to first fix the unknown resonance couplings. The high-energy behaviors of the various form factors and Green functions as dictated by QCD,  can impose strong constraints to the resonance couplings in the R$\chi$T~\cite{Ecker:1989yg,RuizFemenia:2003hm,Cirigliano:2006hb,Guo:2010dv,Chen:2012vw,Roig:2013baa,Chen:2014yta}. Many of the resonance couplings $c_i$ and $d_j$ in Eqs.~\eqref{eq.lagvjp} and \eqref{eq.lagvvp} have been determined in such a way in Refs.~\cite{RuizFemenia:2003hm,Chen:2012vw,Chen:2014yta}. The relevant ones to our study read 
\begin{eqnarray}\label{eq.hecidj}
c_1+4c_3=0\,,\qquad c_1-c_2+c_5=0\,, \qquad c_5-c_6=\frac{N_C M_V}{64 \sqrt2 \pi^2 F_V} \nonumber \\
d_1+8d_2=-\frac{N_c M_V^2}{(8\pi F_V)^2}+\frac{F^2}{4F_V^2},\qquad 
d_3=-\frac{N_c M_V^2}{(8\pi F_V)^2}+\frac{F^2}{8F_V^2}\,,
\end{eqnarray}
with $N_C=3,\,F=F_\pi=0.0924~{\rm GeV},\, M_V=0.775~{\rm GeV}$. The high energy constraints on the $F_A$, $F_V$ and $G_V$ are subject to the resonance operators included in the amplitudes and the ones used in Ref.~\cite{Cirigliano:2002pv} read 
\begin{eqnarray}\label{eq.fvgv2}
F_A=F_\pi,\quad F_V=\sqrt{2} F_\pi, \quad G_V= F_\pi /\sqrt{2} \,. 
\end{eqnarray} 
The large $N_C$ study of the partial-wave $\pi\pi$ scattering by including the crossed-channel effects gives an updated KSRF relation $G_V=F_\pi/\sqrt{3}$~\cite{Guo:2007ff,Guo:2011pa}. Based on the revised KSRF relation, alternative high energy relations can be obtained~\cite{Guo:2010dv,Roig:2013baa,Miranda:2020wdg} 
\begin{eqnarray}\label{eq.fvgv3}
F_A=\sqrt{2} F_\pi,\quad F_V=\sqrt{3} F_\pi, \quad G_V= F_\pi /\sqrt{3} \,.
\end{eqnarray}
We will take the two sets of high energy constraints in Eqs.~\eqref{eq.fvgv2} and \eqref{eq.fvgv3} in later study and the uncertainties of such constraints can be considered as compensation of neglecting the higher order terms in the resonance chiral Lagrangian.  
By taking the above constraints, we are still left with several free parameters that will be fixed by taking the on-shell approximations of the transition vertex involving $\omega\pi$ states. After this procedure, the only remaining unknown parameter is $d_4$ in Eq.~\eqref{eq.lagvvp}, and it is found that the $\omega\to\pi\pi\gamma$ decay width can be used to estimate its value. We find two solutions for $d_4$: a negative value of $d_4=-0.12\pm 0.05$ and a positive one with $d_4=0.82\pm 0.05$. The positive solution with larger magnitude does not fall in the ballpark estimation of the $d_4$ in Ref.~\cite{Miranda:2020wdg}. The detailed discussions can be found in the Appendix~\ref{sec.ff}. In the following numerical discussions, we will take four different combinations of the input resonance couplings, and the deviations resulting from the four situations can be considered as the theoretical uncertainties. In the case of Our-1A, we use the high energy constraints in Eq.~\eqref{eq.fvgv2} and take $d_4=-0.12$. In the case of Our-2A, we take the same value of $d_4=-0.12$ as in the situation of Our-1A, but use the alternative constraints in Eq.~\eqref{eq.fvgv3}. For the cases of Our-1B and Our-2B, the same value of $d_4=0.82$ will be employed and the high energy constraints will be taken from Eq.~\eqref{eq.fvgv2} and Eq.~\eqref{eq.fvgv3}, respectively.

In the $\tau^-\to\pi^-\pi^0\gamma\nu_\tau$ process, there are several types of differential decay widths that are worth in-depth consideration,  since they can reveal different kinds of resonance interactions. For example, the invariant-mass distribution of the $\pi\pi$ system allows one to probe the nonradiative form factor $F_V(t)$, which is dominated by the strong interaction of the $P$-wave $\pi\pi$ system and the $\rho$ resonance. The differential widths of the $\pi^0\gamma$ and $\pi^-\gamma$ provide important environments to study the radiative decay mechanisms of the $\omega$ and $\rho^-$ resonances, respectively. The amplitude of Eq.~\eqref{eq.defT} in our study contains infrared divergence that is caused by the soft photons. A nonvanishing photon-energy cutoff needs to be introduced, which is also usually required by the experimental measurement due to the finite energy resolution of the photon detection. The differential decay width as a function of the photon energy $E_\gamma$ allows us to study the sensitivities of the branching ratios to $E_\gamma$, which can provide important guides for the experimental measurement to implement the proper energy cuts of the photons. In this work, although we use the RAMBO generator~\cite{Kleiss:1985gy} to handle the kinematics, the general discussions on the four-body phase space are also provided in the Appendix~\ref{sec.phasespace}.

\begin{figure}[htbp]
\centering
\includegraphics[width=0.8\textwidth]{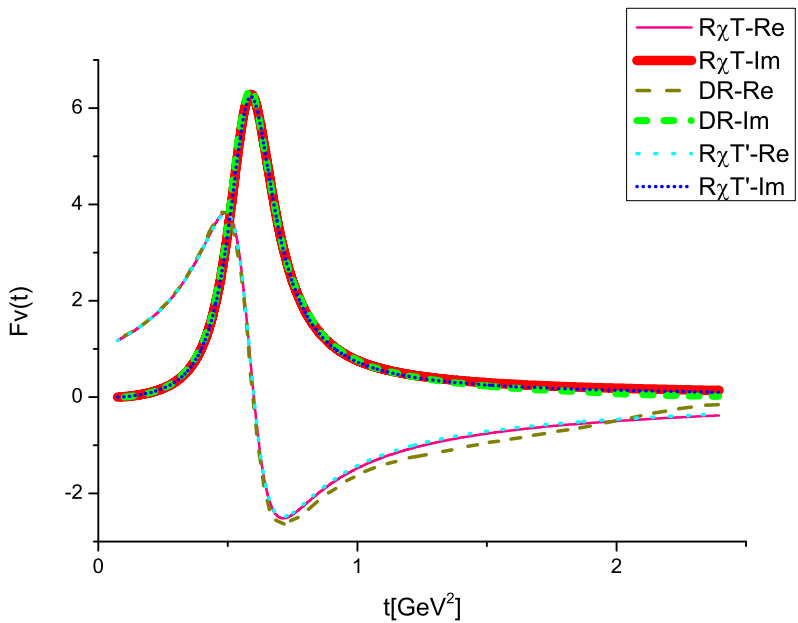}\\
\caption{ \newtex{ The real (Re) and imaginary (Im) parts of the pion vector form factor $F_V(t)$. The curves labeled as R$\chi$T are taken from the full amplitudes from Ref.~\cite{Guerrero:1997ku}, while the symbol R$\chi$T' refers to removing the two-kaon channel from the full amplitudes. The dispersive results of Ref.~\cite{GomezDumm:2013sib} are denoted as $DR$.  }
}\label{fig.fvreim}
\end{figure}

\newtex{Another important input in our study is the pion vector form factor appearing in Eqs.~\eqref{eq.Fnu} and \eqref{eq.Vmunu}. To make direct comparisons with Ref.~\cite{Cirigliano:2002pv}, we use the same pion vector form factor as employed in the former reference, which is originally calculated within the resonance chiral theory by taking into account the constraints of analyticity and unitarity in Ref.~\cite{Guerrero:1997ku}. The explicit expression for $F_V(t)$ reads
\begin{eqnarray}\label{eq.fvgp}
F_V(t) = M_\rho^2 D_\rho  \exp\bigg\{ \frac{-t}{96\pi^2F_\pi^2} \mathcal{R}e \big[ B(t,m_\pi^2) + \frac{1}{2} B(t,m_K^2)  \big]  \bigg\}\,,
\end{eqnarray}
where the loop function $B(t,m_P^2)$ is given by 
\begin{eqnarray}
B(t,m_P^2)= \ln\bigg( \frac{m_P^2}{\mu^2} \bigg) + \frac{8m_P^2}{t} - \frac{5}{3} + \sigma_P^3 \ln\bigg( \frac{\sigma_P+1}{\sigma_P-1} \bigg) \,,
\end{eqnarray}
with
$\sigma_P=\sqrt{1-\frac{4m_P^2}{t}}$. The finite-width effects are included in the $\rho(770)$ propagator $D_\rho$, which explicit expression is given in Eqs.~\eqref{eq.defprogd} and \eqref{eq.widthrho}. As usual, the renormalization scale $\mu$ is set at  the mass of the $\rho(770)$. We further give the result in Fig.~\ref{fig.fvreim} by neglecting the contribution from the two-kaon channel of Eq.~\eqref{eq.fvgp}. It is clear that the two-kaon contribution is tiny. We have also tried to use the recent dispersive result of the pion vector form factor of Ref.~\cite{GomezDumm:2013sib}, and the two form factors turn out to be rather similar, see the curves in Fig.~\ref{fig.fvreim}. As a result, the corresponding figures and numbers by using the dispersive form factor are similar as those by taking the form factor in Eq.~\eqref{eq.fvgp}. Therefore, we refrain from discussing the outputs by using the dispersive form factor in the following. }

\begin{figure}[htbp]
\centering
\includegraphics[width=0.85\textwidth]{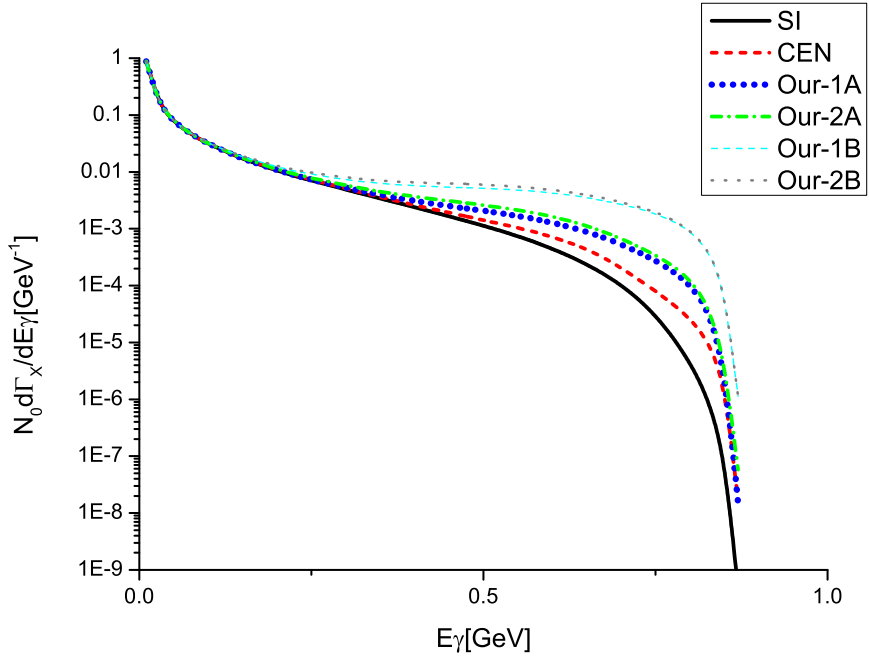}\\
\caption{  Photon energy distribution in the $\tau^-\to\pi^-\pi^0\gamma\nu_\tau$ process. Notice that the differential width of $\tau^-\to\pi^-\pi^0\gamma\nu_\tau$ is normalized to the decay width of the non-radiative process $\tau^-\to\pi^-\pi^0\nu_\tau$, i.e. $N_0= 1/ \Gamma_{\tau^-\to\pi^-\pi^0\nu_\tau}$. The curve labeled as SI is the result by only taking the inner bremsstrahlung contributions. The curve of CEN stands for the result by taking the same amplitude of Ref.~\cite{Cirigliano:2002pv}. And our results, labeled as Our-1A, Our-2A, Our-1B and Our-2B, are obtained by using variant resonance couplings as the phenomenological inputs. See the text for details about the meanings of different notations. 
}\label{fig.eg}
\end{figure}

The dependences of the differential decay widths on the photon energy $E_\gamma$ in the $\tau$ rest frame are given in Fig.~\ref{fig.eg}, where one can clearly see the infrared divergence behavior when $E_\gamma$ approaches to zero. When increasing the photon energy cutoffs, the  branching ratios would rapidly decrease, according to Fig.~\ref{fig.eg}. In Table~\ref{tab.widtheg}, we give the explicit values of the branching ratios of the $\tau^-\to\pi^-\pi^0\gamma\nu_\tau$ process obtained at several intermediate photon energy cutoffs for all the four scenarios with different resonance parameters as inputs.

\begin{table}[htbp]
\centering
\begin{scriptsize}
\begin{tabular}{ c c c c c c c }
\hline\hline
$E_\gamma^{\rm cut}$ & SI   & CEN & Our-1A & Our-2A & Our-1B & Our-2B \\
			\hline
			100MeV & $7.9$ & $8.3$ & $8.7/9.6/8.6/9.4$ & $9.5/10/9.2/9.7$ & $13/9.6/12/9.4$ & $14/10/13/9.7$\\
			\hline
			300MeV & $1.5$ & $1.8$ & $2.4/3.0/2.3/2.8$ & $2.9/3.3/2.6/3.0$ & $5.6/3.0/5.2/2.8$ & $6.3/3.3/5.5/3.0$\\
			\hline
			500MeV & $0.26$ & $0.40$ & $0.73/1.0/0.68/0.90$ & $0.93/1.1/0.81/0.91$& $2.6/1.0/2.4/0.90$ & $2.9/1.1/2.4/0.91$ \\
			\hline\hline
\end{tabular}
\end{scriptsize}
\caption{Branching ratios of the $\tau^-\to\pi^-\pi^0\gamma\nu_\tau$ by taking different photon energy cuts $E_\gamma^{\rm cut}$. All the entries in each column (except the first one) are multiplied by $10^{-4}$. The results in the columns labeled by SI and CEN are obtained by taking the structure independent expressions and the full amplitudes from~Ref.~\cite{Cirigliano:2002pv}, respectively. The tiny differences between the numbers of Ref.~\cite{Cirigliano:2002pv} and ours are caused by using the updated physical constants, especially the pion decay constant $F_\pi$. The meanings of the notations of Our-1A, Our-2A, Our-1B, Our-2B are explained in the text. For each entry in the last four columns, there are four numbers and they  correspond to taking different contributions in the decay amplitudes. The first number is for taking the full amplitudes in our study, the second one is obtained by omitting the $\omega\rho\pi$ interaction vertices, i.e. Feynman diagrams of $V_a$ and $V_c$ in Fig.~\ref{fig.v}, the third one is to exclude the contribution from the axial-vector resonance, i.e. the $F_A$ term of Eq.~\eqref{eq.laga2}, and the fourth number is derived by neglecting the contributions from both the $\omega\rho\pi$ vertices and the axial-vector resonance.   \label{tab.widtheg}   } 
\end{table}

Let's first analyze the results from the scenarios of Our-1A and Our-2A by taking $d_4=-0.12$. The full results at $E_\gamma^{\rm cut}= 100$ and 300~MeV in these two scenarios are close to the branching ratios predicted in Ref.~\cite{Cirigliano:2002pv}, which are explicitly given in the column labeled as CEN in Table~\ref{tab.widtheg}. While, our predictions in these two scenarios are different from the result of Ref.~\cite{Cirigliano:2002pv} when taking larger photon energy cutoff at $E_\gamma^{\rm cut}= 500$~MeV. This is within the expectation, since the total width of the $\tau^-\to\pi^-\pi^0\gamma\nu_\tau$ process is overwhelmingly dominated by the model-independent inner bremsstrahlung part when taking small photon energy cutoffs $E_\gamma^{\rm cut}$. The SD contributions in Eqs.~\eqref{eq.Vmunu} and \eqref{eq.Amunu} would play more important roles in the $\tau^-\to\pi^-\pi^0\gamma\nu_\tau$ decay when taking larger photon energy cutoffs. Precisely, the key differences between our study and the one in Ref.~\cite{Cirigliano:2002pv} are the SD parts, namely the $v_{i=1,2,3,4}$ and $a_{i=1,2,3,4}$ form factors in Eqs.~\eqref{eq.Vmunu} and \eqref{eq.Amunu}. This also indicates that it is helpful to distinguish different hadronic models in the $\tau^-\to\pi^-\pi^0\gamma\nu_\tau$ process by taking larger photon energy cutoffs, although its branching ratio will be reduced.

\begin{figure}[htbp]
\centering
\includegraphics[width=0.83\textwidth]{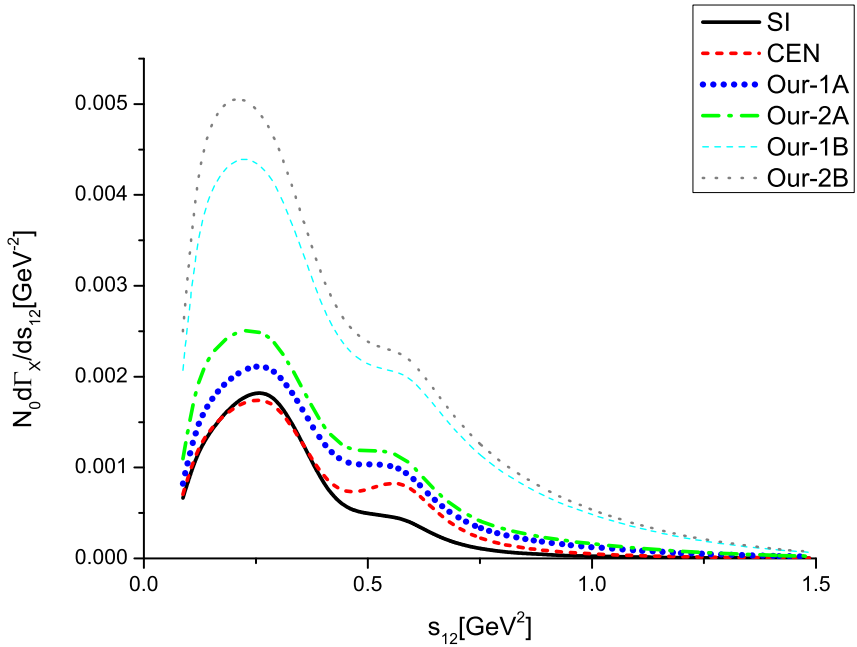}
\caption{Invariant-mass distributions of the $\pi\pi$ with $E_{\gamma }^{\rm cut}=0.3$~GeV. The normalization factor is taken as $N_0=1/ \Gamma_{\tau^-\to\pi^-\pi^0\nu_\tau}$. The meanings of different curves are the same as those in Fig.\ref{fig.eg}. 
}\label{fig.pipi}
\end{figure}

In addition, we also further investigate the specific roles played by the intermediate $\omega$ and $a_1$ resonances. We distinguish three different situations by separately excluding the effect of the $\omega\rho\pi$ vertices and the $a_1$ effect, and omitting both of these two contributions. The last three numbers of each entry in the last four columns of Table~\ref{tab.widtheg} are derived under the aforementioned three situations, respectively. We only observe mild changes of the four numbers in each entry for the columns Our-1A and Our-2A, indicating that the $\omega\rho\pi$ interactions and the $a_1$ resonance do not seem to play the decisive roles in the $\tau\to \pi\pi\gamma\nu_\tau$.   

For the scenarios of Our-1B and Our-2B by using $d_4=0.82$, the branching ratios from the full amplitudes, look larger than the results from the scenarios of Our-1A and Our-2A by taking $d_4=-0.12$. Nevertheless, for a given photon energy cutoff, the orders of magnitudes from different scenarios remain the same. According to the Belle-II estimate~\cite{Belle-II:2018jsg}, around 45 billion pairs of the $\tau$ leptons will be collected. Hence we anticipate that the radiative two-pion $\tau$ decay processes have the good chance to be measured by the Belle-II experiment. 

Next we discuss several interesting invariant-mass spectra of different two-particle systems. To be definite, all the two-particle spectra shown below are plotted by taking the cutoff $E_\gamma^{\rm cut}=300$~MeV. 
The invariant-mass distributions of the $\pi\pi$ system are shown in Fig.~\ref{fig.pipi}. We give several different curves in this figure, and the meanings of the labels for the curves are the same as those in Table~\ref{tab.widtheg}. Comparing with the differences between the SI case and other situations, one can conclude that the bump of the $\rho$ resonance is obviously enhanced in the $\pi\pi$ spectrum when the SD form factors $v_{i=1,2,3,4}$ are included. The similarity between the curves of the CEN and Our-1A and Our-2A around the $\rho$ energy region indicates that the inclusion of the odd-intrinsic parity operators in Eqs.~\eqref{eq.lagvvp} and \eqref{eq.lagvjp} mildly affects the $\pi\pi$ spectrum in that energy region with $d_4=-0.12$. Nevertheless, the heights of the curves corresponding to the cases of Our-1B and Our-2B by taking $d_4=0.82$ are clearly larger than those of the other cases. \newtex{It should be noted that the positive value of $d_4$ with large magnitude is not within the guesstimate range in Ref.~\cite{Miranda:2020wdg}. In addition, the unexpected huge effects from the $VVP$ interactions in the scenarios  Our-1B and Our-2B, with respect to the minimal resonance chiral operators labeled by the CEN curves, give another hint that the large magnitude of $d_4$ does not seem to be the physical solution. The same conclusions are also applied to the following discussions about the distributions of the $\pi\gamma$ and $A_\xi$.}

\begin{figure}[htbp]
\centering
\includegraphics[width=0.49\textwidth]{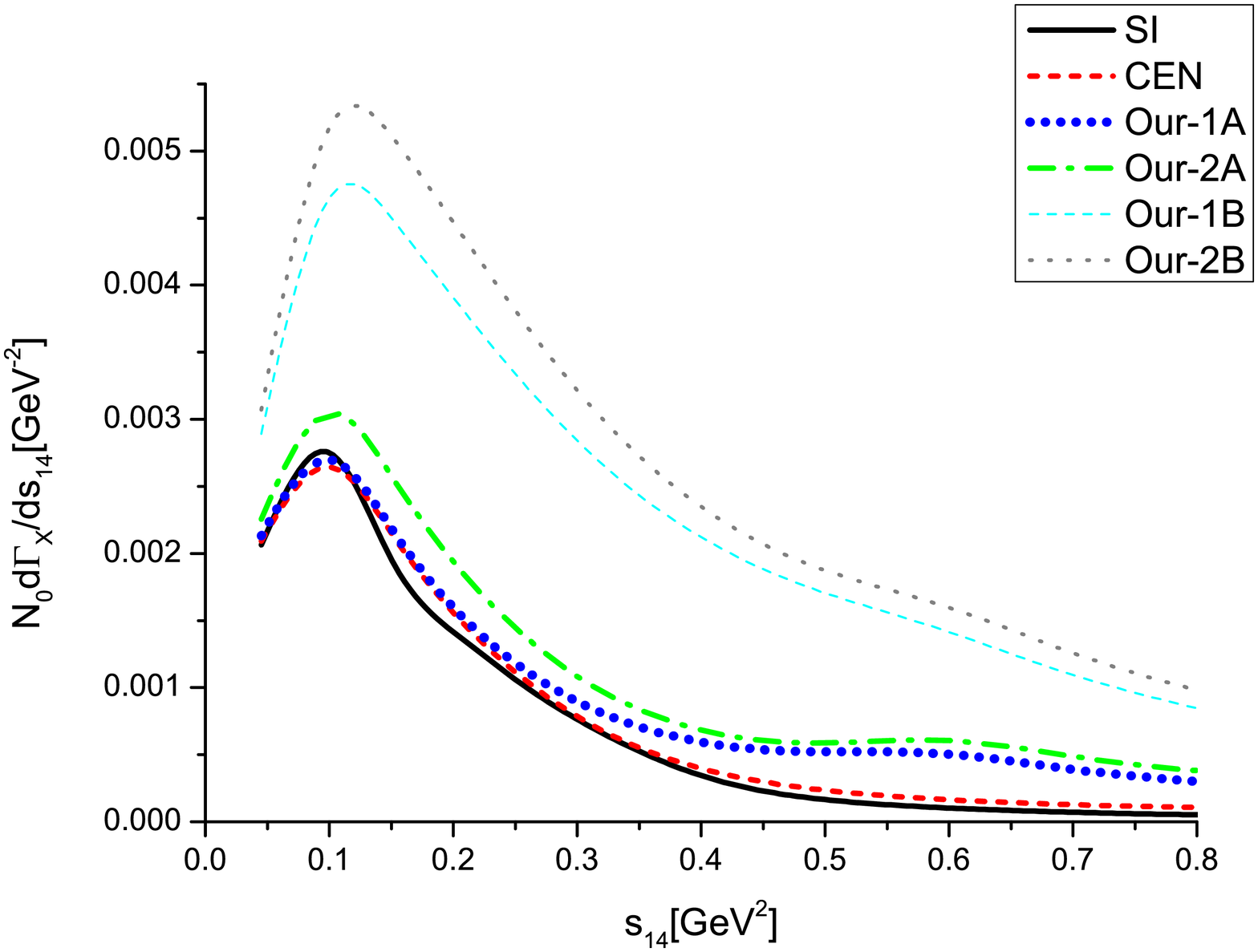}
\includegraphics[width=0.49\textwidth]{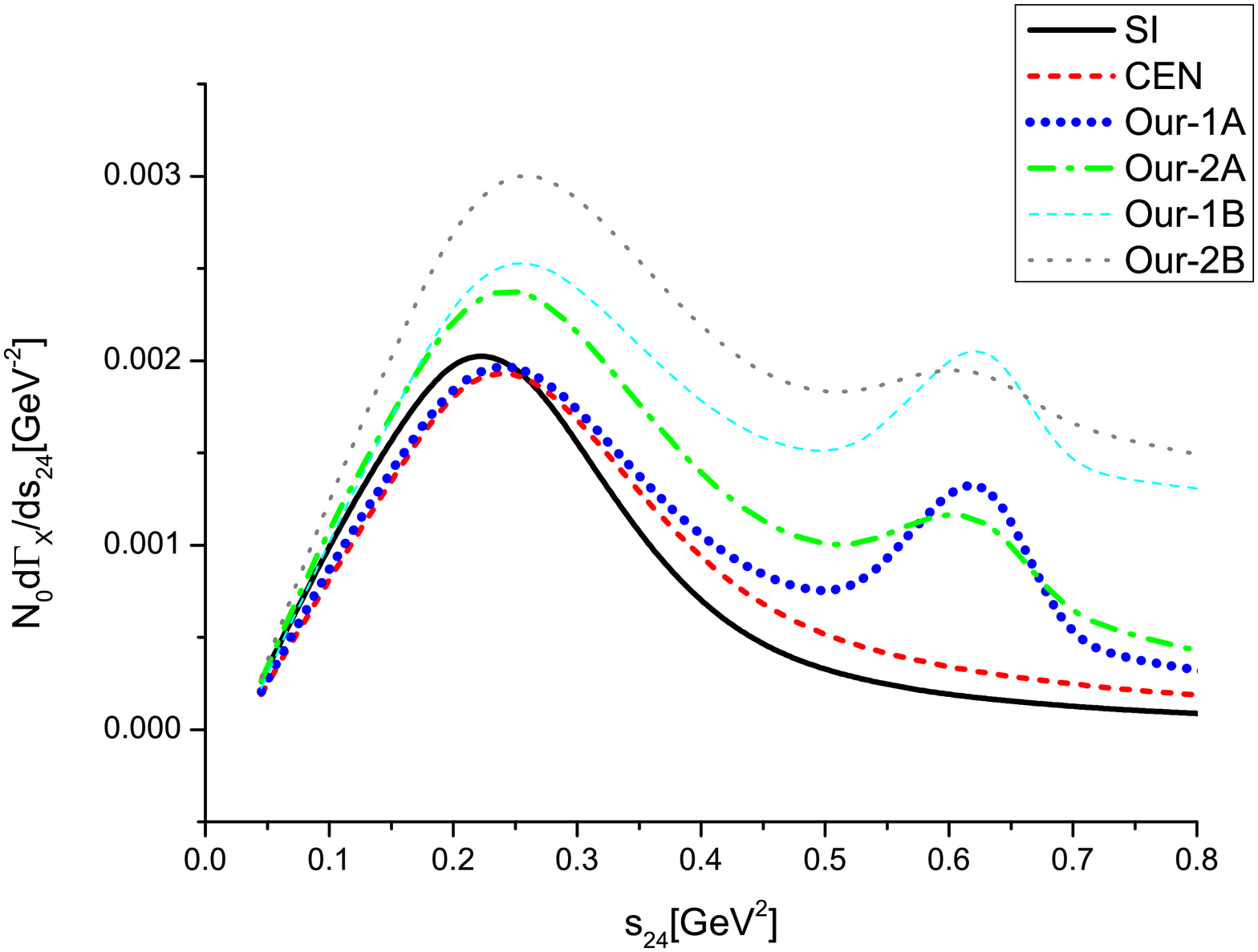}
\caption{Differential decay widths of the $\pi^-\gamma$ (left) and the $\pi^0\gamma$ (right) with $E_{\gamma}^{\rm cut}=0.3$~GeV. The normalization factor $N_0$ is the inverse of the decay width of $\tau^-\to\pi^-\pi^0\nu_\tau$, i.e. $N_0= 1/ \Gamma_{\tau^-\to\pi^-\pi^0\nu_\tau} $. The meanings of different curves are the same as those in Fig.\ref{fig.eg}. 
 }\label{fig.pig}
\end{figure}

\begin{figure}[htbp]
	\centering
	\includegraphics[width=0.82\textwidth]{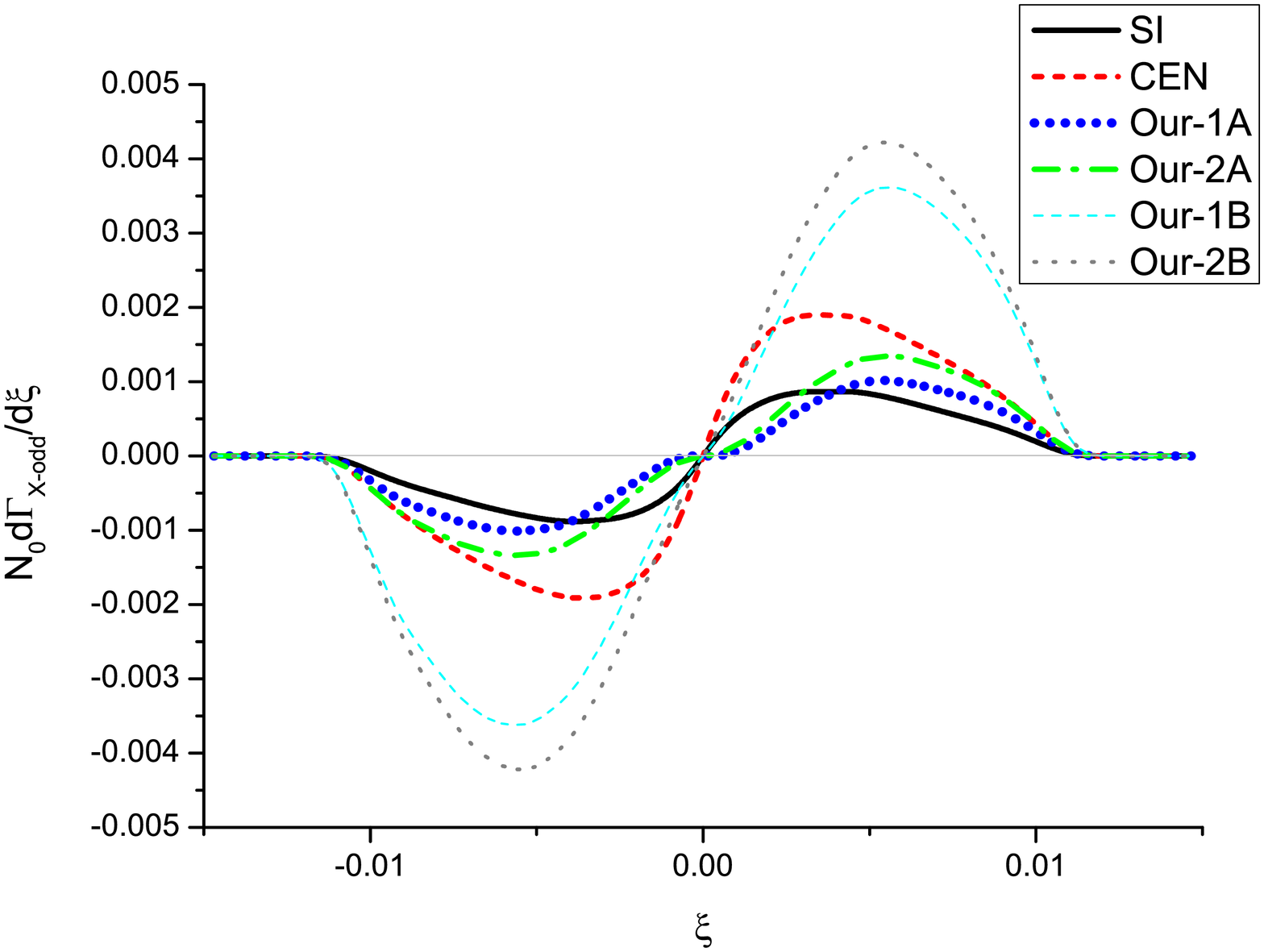}\\
	\caption{Distributions of the triple-product asymmetry $A_\xi$ with $E_{\gamma }^{\rm cut}=0.3$~GeV. The normalization factor is taken as $N_0=1/ \Gamma_{\tau^-\to\pi^-\pi^0\nu_\tau}$. The meanings of different curves are the same as those in Fig.\ref{fig.eg}. 
	}\label{fig.xiall}
\end{figure}

The invariant-mass spectra of the $\pi^-\gamma$ and $\pi^0\gamma$ systems are given in the left and right panels of Fig.~\ref{fig.pig}, respectively. According to the Feynman diagrams in Figs.~\ref{fig.v} and \ref{fig.a}, both the narrow $\omega$ and broad $\rho^0$ contribute to the $\pi^0\gamma$ distribution, while only the broad $\rho^-$ resonance enters in the $\pi^-\gamma$ spectrum. The different resonant contents in the charged $\pi^-\gamma$ and neutral $\pi^0\gamma$ channels are clearly reflected in their invariant-mass distributions, as shown in Fig.~\ref{fig.pig}. Furthermore, the resonance effects enter the $\pi\gamma$ spectra via the $VVP$ and $VJP$ types of operators, which are absent in the SI and CEN amplitudes. This also explains the smooth curves  from the latter two cases in Fig.~\ref{fig.pig}. Similar as the situation in the two-pion distribution in Fig.~\ref{fig.pipi}, the heights of the curves from the scenarios of Our-1B and Our-2B are larger than those of other cases. A future experimental measurement on the charged and neutral $\pi\gamma$ distributions will definitely be useful to constrain the anomalous interactions of the $\rho$ and $\omega$ resonances.

\begin{figure}[htbp]
	\centering
	\includegraphics[width=0.43\textwidth]{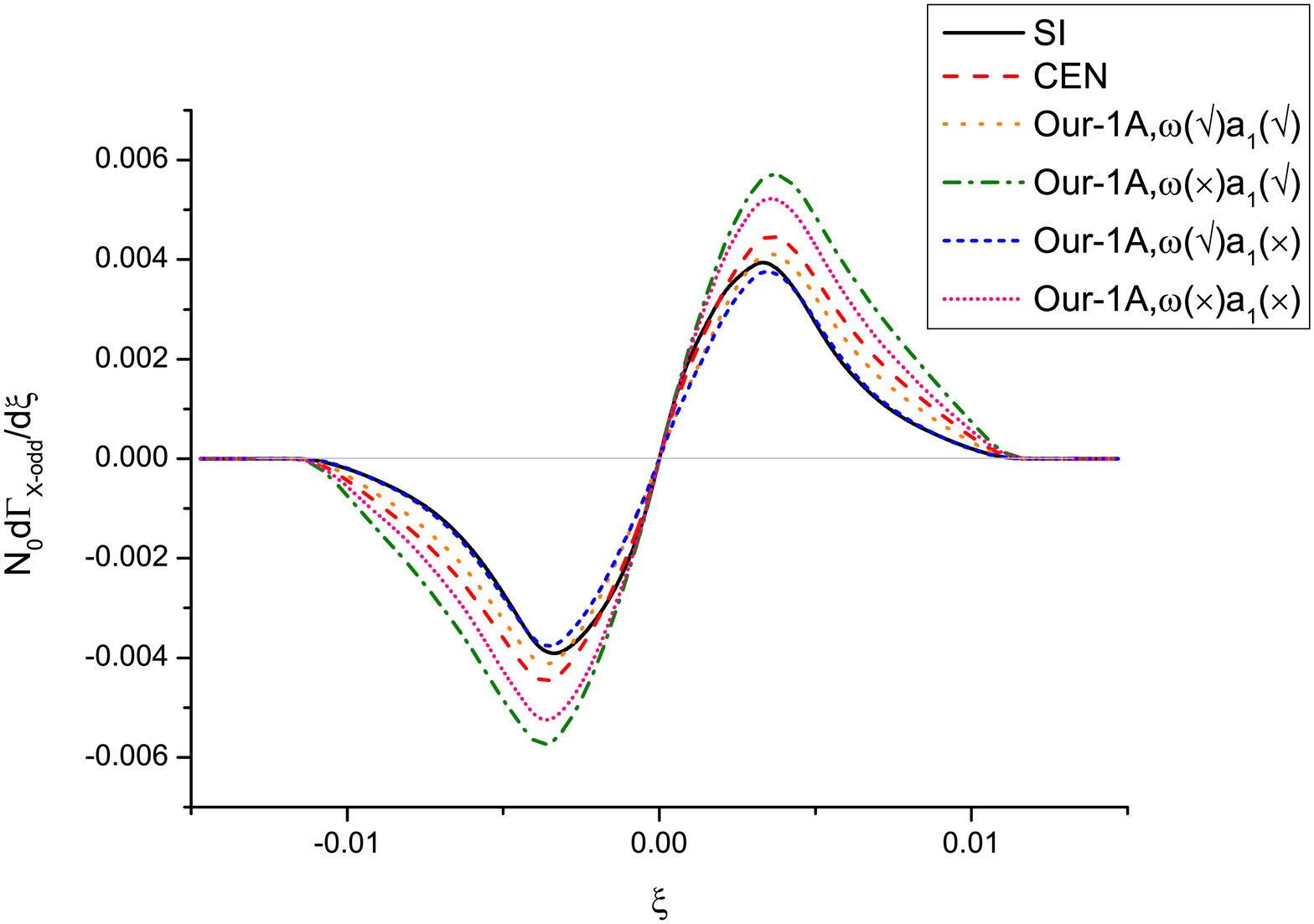}
	\includegraphics[width=0.43\textwidth]{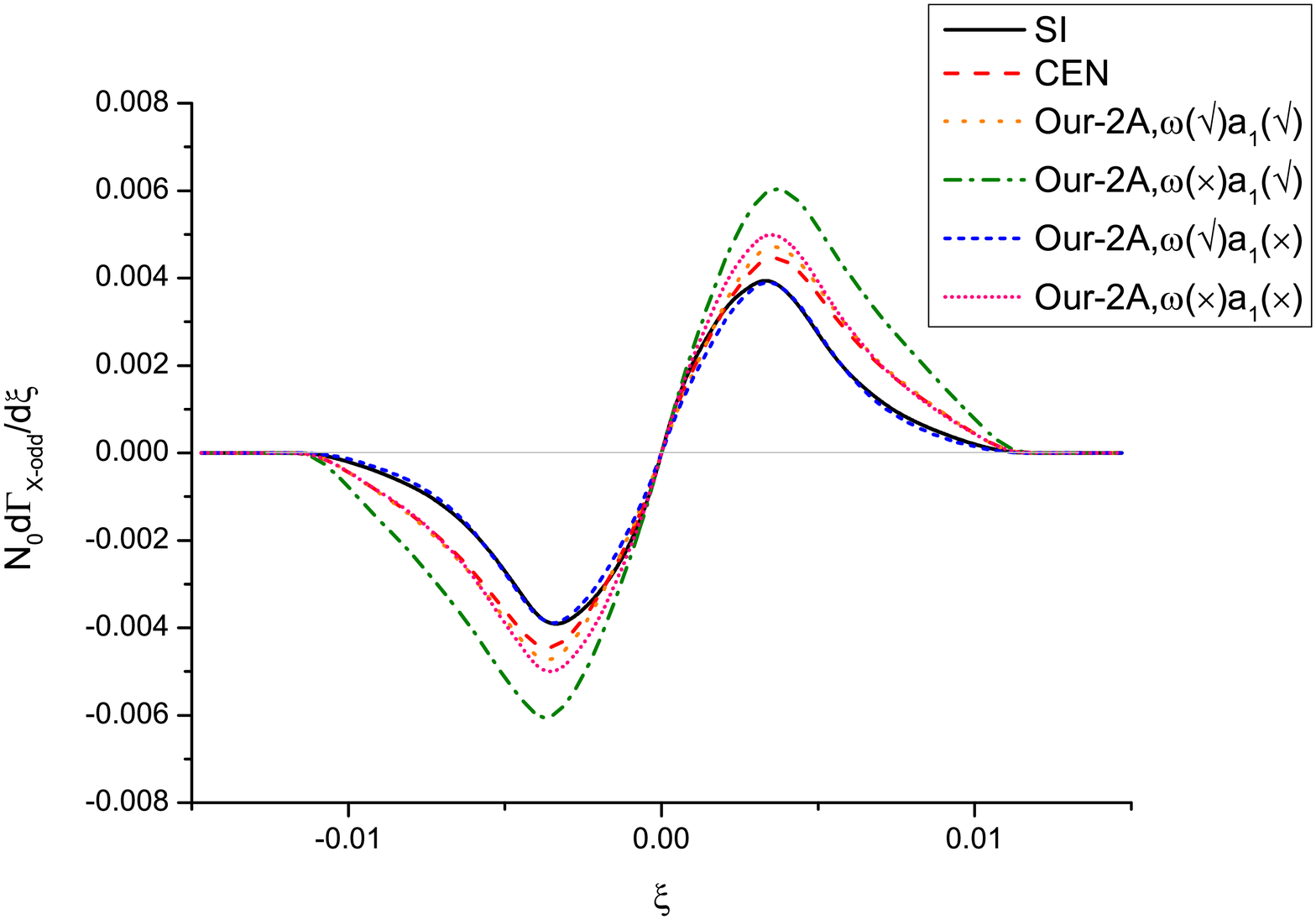}
	\includegraphics[width=0.43\textwidth]{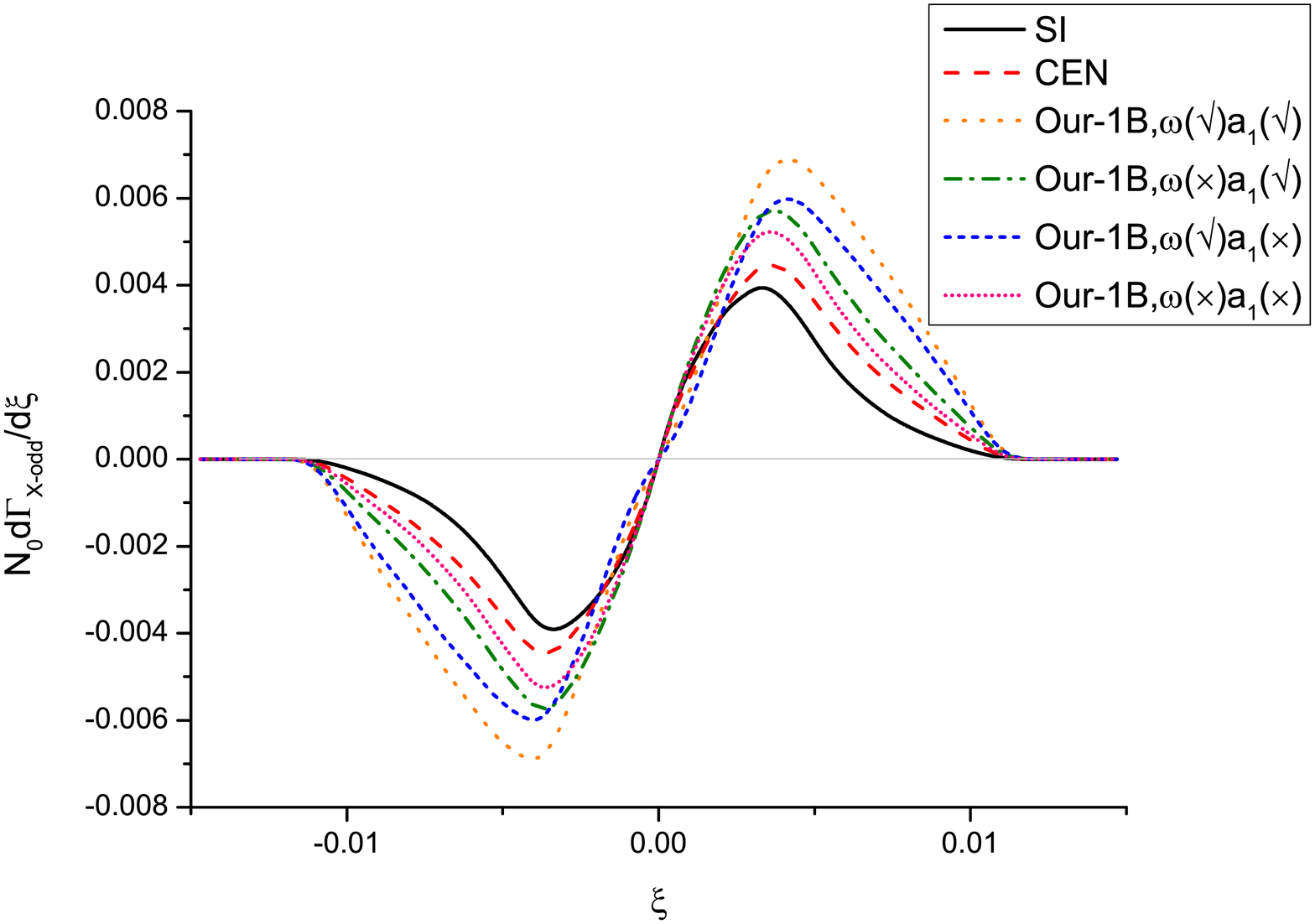}
	\includegraphics[width=0.43\textwidth]{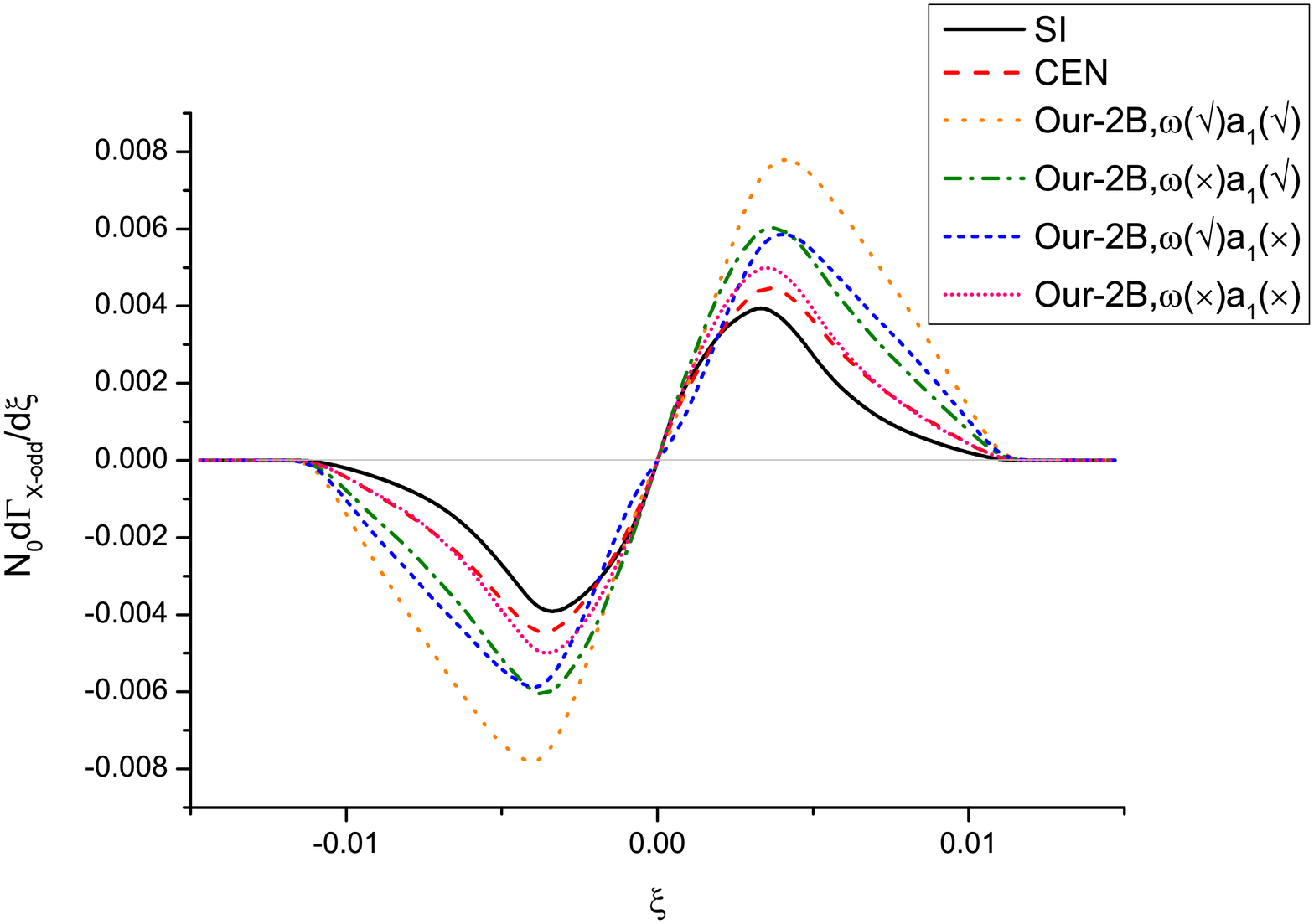}
	\caption{ Comparisons of the $\xi$ distributions by excluding($\times$)/including($\surd$) the $\omega\rho\pi$ interactions and the $a_1$ resonance. The notations of other symbols are the same as those in Fig.~\ref{fig.eg}.  
	}\label{fig.xi1}
\end{figure}

The triple-product asymmetry distributions with respect to the $\xi$ are shown in Fig.~\ref{fig.xiall}. The first lesson we can learn from Fig.~\ref{fig.xiall} is that sizable nonvanishing triple-product asymmetry distributions are obtained regardless of taking which amplitudes among the SI, CEN and ours. Nevertheless, it is also clear that different amplitudes or couplings can lead to obviously different curves. On the other hand, this implies that the measurement of the $\xi$ distributions can definitely provide new experimental criteria to discern different hadronic models. To further study the roles of different resonance interactions in the $\xi$ distributions, we distinguish several different situations by excluding/including the effects of the $\omega\rho\pi$ interacting vertices and the $a_1$ resonance in Fig.~\ref{fig.xi1}, so that one could discern the relative strengths of the vector (specially the $\omega\rho\pi$ interactions) and axial-vector resonances. According to the curves in Fig.~\ref{fig.xi1}, loosely speaking the roles of $\omega\rho\pi$ interactions played in the $\xi$ distributions seem similar as those of the broad $a_1$ resonance. Comparing the curves resulting from separately excluding the $\omega\rho\pi$ vertices and the $a_1$ with the full results, though in some cases, such as in Our-1A, the $\xi$ distribution curves are slightly more influenced by the $\omega\rho\pi$ interacting vertices than the $a_1$. Nevertheless, the roles of different resonance interactions are also affected by the input parameters.

\begin{table}[htbp]
	\centering
	\begin{scriptsize}
		\begin{tabular}{ c c c c c }
			\hline\hline
			$E_\gamma^{\rm cut}$  & $A_\xi$(Our-1A)        & $A_\xi$(Our-2A)  & $A_\xi$(Our-1B)        & $A_\xi$(Our-2B)      \\
			\hline
			100~MeV            & $1.2/1.7/1.0/1.6$ & $1.3/1.8/0.98/1.4$ & $1.6/1.7/1.4/1.6$ & $1.7/1.8/1.3/1.4$\\
			\hline
			300~MeV            & $1.5/2.6/1.0/2.2$ & $1.6/2.5/0.73/1.6$ & $2.3/2.6/2.0/2.2$ & $2.4/2.5/1.7/1.6$\\
			\hline
			500~MeV            & $0.98/1.4/0.58/0.88$ & $0.91/1.4/0.68/0.43$ & $2.1/1.4/1.8/8.8$ & $2.1/1.4/1.5/4.3$\\ 
			\hline\hline
		\end{tabular}
	\end{scriptsize}
	\caption{Rates of the triple-product asymmetry with different photon energy cutoffs. All the numbers except those in the first column are multiplied  by the factor of $10^{-2}$. The meanings of different notations and the way to obtain the four numbers in each entry in the last four columns are the same as those in Table~\ref{tab.widtheg}. 
	} \label{tab.asymmetry} 
\end{table}

In Table~\ref{tab.asymmetry}, we give the integrated rates of the asymmetries shown in Fig.~\ref{fig.xiall}, i.e., the values of $A_\xi$. Comparing with the  values of $A_\xi\sim 10^{-4}$ for the $K_{l3\gamma}$ decays~\cite{Braguta:2001nz,Rudenko:2011qe,Muller:2006gu}, the corresponding results of the triple-product asymmetries in the $\tau\to\pi\pi\gamma\nu_\tau$ are around two orders of magnitudes larger, and hence are very encouraging for the future experimental measurements conducted in Belle-II~\cite{Belle-II:2018jsg} and the prospective super tau-charm facility~\cite{bib.stcf}. Moreover, different amplitudes with different resonance couplings lead to obviously different shapes of the triple-product asymmetry distributions with respect to the $\xi$ variable. Therefore the precise measurements of the triple-product asymmetry distributions as functions of $\xi$ also provide another quantity to discern different resonance dynamics in the $\tau\to\pi\pi\gamma\nu_\tau$ process.

\section{Summary and conclusions}\label{sec.conclusion}

In this work, we perform a scrutinized study of the $\tau^-\to\pi^-\pi^0\gamma\nu_\tau$ process within the resonance chiral theory. In addition to the minimal resonance chiral Lagrangian, the odd-intrinsic parity resonance operators are also included in our calculation. The unknown resonance couplings are completely fixed by taking the high energy constraints, the on-shell approximations of the $\omega\pi$ transition vertices and the $\omega\to\pi\pi\gamma$ decay width. This allows us to give pure predictions to the various observables in the $\tau^-\to\pi^-\pi^0\gamma\nu_\tau$ process. By taking the photon energy cutoffs in the range between 100~MeV and 500~MeV, the decay branching ratios of the $\tau^-\to\pi^-\pi^0\gamma\nu_\tau$ process are estimated to vary from around $0.7\times 10^{-4}$ to $1.4\times 10^{-3}$. We prospect that the $\tau^-\to\pi^-\pi^0\gamma\nu_\tau$ has the good chance to be measured in the Belle-II and the future super-charm factory. 

The prominent bump around the $\rho$ resonance energy region is predicted in the invariant-mass distribution of the $\pi\pi$ system. The significant peak is expected to show up in $\pi^0\gamma$ spectrum, due to the contribution of the narrow $\omega$. In contrast, the $\pi^-\gamma$ spectrum turns out to be smooth. The most important finding of this work is the promising nonzero triple-product asymmetry distribution. It is encouraging that the Belle-II collaboration and the prospective super tau-charm facilities may be able to observe this intriguing asymmetry distribution in the $\tau^-\to\pi^-\pi^0\gamma\nu_\tau$ process.

\section*{Acknowledgements}
We would like to thank J.~Portol\'es and P.~Roig for useful discussions. We also thank B.~Kubis for communications. This work is partially funded by the Natural Science Foundation of China (NSFC) under Grant Nos.~11975090, 12150013 and 11575052, and the Fundamental Research Funds for the Central Universities. 

\appendix

\section{Four-body decay kinematics}\label{sec.phasespace}
\setcounter{equation}{0}
\def\theequation{\Alph{section}.\arabic{equation}}

Here we follow the method introduced in Refs.~\cite{Axelrod:1984,Nyborg:1965zz,Kumar:1969jjy} to treat the phase space integrals. In this method, 
all the variables appearing in the phase space integrals are Lorentz scalars. 
For a process $p \to p_1 + p_2 + p_3 + p_4$, the corresponding phase space integral is
\begin{equation}\label{eq.defps}
\begin{aligned}
\Phi_{4} &=\frac{1}{(2\pi)^{12}}\int \frac{d^{3} \overrightarrow{p_{1}}}{2 E_{1}} \frac{d^{3} \overrightarrow{p_{2}}}{2 E_{2}} \frac{d^{3} \overrightarrow{p_{3}}}{2 E_{3}} \frac{d^{3} \overrightarrow{p_{4}}}{2 E_{4}} \delta^{(4)}\left(p-p_{1}-p_{2}-p_{3}-p_{4}\right) \\
&=\frac{\pi^{2}}{32(2\pi)^{12} m_{0}^{2}} \int d S_{12} d S_{123} d S_{34} d S_{13} d S_{134} \frac{1}{\sqrt{-\Delta_{4}\left(p_{1}, p_{2}, p_{3}, p_{4}\right)}}\,, 
\end{aligned}
\end{equation}
where
\begin{equation}
p^{2}=m_{0}^{2}, p_{1}^{2}=m_{1}^{2}, p_{2}^{2}=m_{2}^{2}, p_{3}^{2}=m_{3}^{2}, p_{4}^{2}=m_{4}^{2}\,,
\end{equation}
\begin{equation}
S_{i j}=\left(p_{i}+p_{j}\right)^{2}, S_{i j k}=\left(p_{i}+p_{j}+p_{k}\right)^{2}\,,
\end{equation}
and $\Delta_{4}$ is the Gram determinant~\cite{Byckling:1971vca}
\begin{equation}
\Delta_{4}\left(p_{1}, p_{2}, p_{3}, p_{4}\right)=\left(\begin{array}{cccc}
p_{1} \cdot p_{1} & p_{1} \cdot p_{2} & p_{1} \cdot p_{3} & p_{1} \cdot p_{4} \\
p_{2} \cdot p_{1} & p_{2} \cdot p_{2} & p_{2} \cdot p_{3} & p_{2} \cdot p_{4} \\
p_{3} \cdot p_{1} & p_{3} \cdot p_{2} & p_{3} \cdot p_{3} & p_{3} \cdot p_{4} \\
p_{4} \cdot p_{1} & p_{4} \cdot p_{2} & p_{4} \cdot p_{3} & p_{4} \cdot p_{4}
\end{array}\right)\,.
\end{equation}
The Gram determinant $\Delta_{4}$ can be written as 
\begin{equation}
-\Delta_{4} \equiv \hat{a} S_{134}^{2}+\hat{b} S_{134}+\hat{c}\,,
\end{equation} 
with 
\begin{equation}\label{eq.delta4coef}
\begin{aligned}
\hat{a}=&-[(S_{123}-S_{12})^2-2(S_{123}+S_{12})m_3^2+m_3^4]/16 \,,\\
\hat{b}=&\{ 
 - S_{13} [m_0^2 (m_3^2 + S_{12} - S_{123}) + S_{12} (m_3^2 - 2 m_4^2 - S_{12} + S_{123})] \\
&- S_{34} [m_2^2 (m_3^2 + S_{123} - S_{12}) + S_{123} (m_3^2 - 2 m_1^2 - S_{123} + S_{12}) ] \\
&+S_{13} S_{34} (m_3^2 - S_{12} - S_{123} )+ S_{12} S_{123} (2 m_3^2 - m_1^2 - m_4^2 ) \\
&+ S_{12} [m_0^2 (m_1^2 - m_3^2) - m_2^2 (m_3^2 + m_4^2) - m_4^2 (m_1^2 + m_3^2 - S_{12})] \\
&+ S_{123} [m_2^2 (m_4^2 - m_3^2) - m_0^2 (m_3^2 + m_1^2) - m_1^2 (m_4^2 + m_3^2 - S_{123})] \\
&+ m_3^2 [m_0^2 (m_3^2 - m_1^2) + m_2^2 (m_3^2 - m_4^2)+ 2 m_0^2  m_2^2 + m_1^2 m_4^2 ] \}/8\,,\\
\hat{b}^{2}-4 \hat{a} \hat{c} =&(\hat{l} S_{13}^2+\hat{m} S_{13}+\hat{n})(\hat{p} S_{34}^2+\hat{q} S_{34}+\hat{r})/16\,,\\
\hat{l}&=S_{12} \,,\\
\hat{m}&=S_{12}^2-(S_{123}+m_3^2+m_2^2+m_1^2)S_{12}-(m_3^2-S_{123})(m_2^2-m_1^2) ,\\
\hat{m}^{2}-4 \hat{l} \hat{n} &=[S_{12}-(m_2-m_1)^2][S_{12}-(m_2+m_1)^2][S_{12}-(m_3-\sqrt{S_{123}})^2][S_{12}-(m_3+\sqrt{S_{123}})^2] \,,\\
\hat{p}&=S_{123} \,,\\
\hat{q}&=S_{123}^2-(S_{12}+m_3^2+m_0^2+m_4^2)S_{123}-(m_3^2-S_{12})(m_0^2-m_4^2) \,,\\
\hat{q}^{2}-4 \hat{p} \hat{r} &=[S_{123}-(m_0-m_4)^2][S_{123}-(m_0+m_4)^2][S_{123}-(m_3-\sqrt{S_{12}})^2][S_{123}-(m_3+\sqrt{S_{12}})^2] \,. 
\end{aligned}
\end{equation}

From the requirements of $-\Delta_{4} \geq 0$ and $\hat{b}^{2}-4 \hat{a}\hat{c} \geq 0$, one can determine the upper and lower limits for the integral variables.
Then one can write the phase space integrals in terms of the Lorentz scalar  variables
\begin{equation}\label{eq.phi4sim}
\Phi_{4}=\frac{\pi^{2}}{32 (2\pi)^{12}m_{0}^{2}} \int_{(m_1+m_2)^2}^{(m_0-m_3-m_4)^2} d S_{12} \int_{(m_3+\sqrt{S_{12}})^2}^{(m_0-m_4)^2} d S_{123} \int_{S_{34}^-}^{S_{34}^+} d S_{34} \int_{S_{13}^-}^{S_{13}^+} d S_{13} \int_{S_{134}^-}^{S_{134}^+}\frac{d S_{134}}{\sqrt{-\Delta_{4}}} \,,
\end{equation}
with
\begin{equation}
\begin{aligned}
S_{134}^\pm &=\frac{-\hat{b} \pm \sqrt{\hat{b}^{2}-4 \hat{a}\hat{c}}}{2 \hat{a}} \,,\\
S_{13}^\pm &=\frac{-\hat{m} \pm \sqrt{\hat{m}^{2}-4 \hat{l}\hat{n}}}{2 \hat{l}}\,,\\
S_{34}^\pm &=\frac{-\hat{q} \pm \sqrt{\hat{q}^{2}-4 \hat{p}\hat{r}}}{2 \hat{p}} \,,
\end{aligned}
\end{equation}
where the definitions of the kinematical variables labeled with hats are given in Eq.~\eqref{eq.delta4coef}.

The above phase-space integral formulas are valid for general situations with arbitrary masses. For the $\tau(p)\to\pi^-(p_1)\pi^0(p_2)\nu_\tau(p_3)\gamma(p_4)$ process, we have 
\begin{equation}
m_{0}^{2}=m_{\tau}^{2}\,, \quad m_{1}^2=m_{2}^2=m_{\pi}^{2}\,, \quad m_{3}^{2}=0\,, \quad m_{4}^{2}=0\,.
\end{equation} 
By taking the changes of variables~\cite{Axelrod:1984}
\begin{equation}
\begin{aligned}
S_{134} &=\frac{1}{2 \hat{a}}\left[-\hat{b}+\sin \left(\widetilde{S_{134}}\right) \sqrt{\hat{b}^{2}-4 \hat{a}\hat{c}}\right]\,, \\
S_{13} &=4 \sqrt{-\hat{a}} \widetilde{S_{13}}+m_{1}^{2}\,,
\end{aligned}
\end{equation}
it is possible to make the $\frac{1}{\sqrt{-\Delta_{4}\left(p_{1}, p_{2}, p_{3}, p_{4}\right)}}$ factor, which is divergent in the integration boundary, absent in Eq.~\eqref{eq.phi4sim}. In this way, the four-body phase-space integral of the $\tau(p)\to\pi^-(p_1)\pi^0(p_2)\nu_\tau(p_3)\gamma(p_4)$ can be cast in a neat form
\begin{equation}
\Phi_{4}=\frac{\pi^{2}}{8(2\pi)^{12}m_{\tau}^{2}} \int_{4 m_\pi^2}^{m_\tau^2} d S_{12} \int_{S_{12}}^{m_\tau^2} d S_{123} \int_{0}^{(S_{123}-S_{12})(m_{\tau}^{2}-S_{123})/S_{123}} d S_{34} \int_{\tilde{S}_{13}^-}^{\tilde{S}_{13}^+} d \widetilde{S_{13}} \int_{-\frac{\pi}{2}}^{\frac{\pi}{2}} d \widetilde{S_{134}}\,,
\end{equation}
with
\begin{equation}
\tilde{S}_{13}^{\pm}=\frac{1}{2}  \pm\sqrt{\frac{1}{4}-\frac{m_\pi^2}{S_{12}}} .
\end{equation}

\section{Form factors and determination of the resonance couplings}\label{sec.ff}
\setcounter{equation}{0}

Although it is straightforward to calculate the relevant Feynman diagrams using the Lagrangians~\eqref{eq.lagv2}-\eqref{eq.lagvjp}, additional free parameters will appear in the $\tau\to\pi\pi\gamma\nu_\tau$, since the high-energy behaviors only impose constraints on some specific combinations of the resonance couplings~\eqref{eq.hecidj}. This motivates us to apply the on-shell Feynman rules to the anomalous $J\omega\pi$ and $J\rho\pi$ contact interacting vertices, being $J$ the charged or neutral vector source fields. This procedure turns out to be very helpful to reduce the unknown resonance couplings in the $\tau\to\pi\pi\gamma\nu_\tau$ decay. 

The on-shell Feynman rule determined from the Lagrangian~\eqref{eq.lagvjp} for the contact $\omega\pi\gamma$ vertex takes the form 
\begin{eqnarray}\label{eq.gw}
g_\omega e \varepsilon_{\mu\nu\rho\sigma} k^\rho
\end{eqnarray}
where the effective coupling $g_\omega$ is given by
\begin{eqnarray}
g_\omega=  \frac{\sqrt2}{ M_V F} \bigg[ (c_2-c_1+c_5-2c_6)M_\omega^2 + (c_1+c_2+8c_3-c_5)m_\pi^2 \bigg] \,.
\end{eqnarray}
According to Ref.~\cite{Ecker:1988te}, the following normalization of the anti-symmetric tensor $\omega_{\mu\nu}$ field
\begin{eqnarray}
<0|\omega_{\mu\nu}|\omega(q)> = \frac{i}{M_\omega}( q_\mu \epsilon_\nu - q_\nu \epsilon_\mu )\,, 
\end{eqnarray}
is used here to derive the on-shell expression of the contact $\omega\pi\gamma$ vertex. The charged anomalous vector current to the $\omega\pi$ transition vertex takes the same form as that in  Eq.~\eqref{eq.gw}, with the obvious replacement of the electric charge by other terms, including the Fermi constant and relevant CKM matrix element, in the definition of the decay amplitude in Eq.~\eqref{eq.defT}. Similarly, for the $\rho\pi\gamma$ vertex in the Feynman diagrams of Fig.~\eqref{fig.a}, we also use the on-shell description and it is related to the $\omega\pi\gamma$ one via the $SU(3)$ flavor symmetry, that is  $g_{\rho}= g_\omega/3$.


With these preparations, we can calculate the Feynman diagrams shown in Figs.~\ref{fig.v} and \ref{fig.a} and then determine the various vector and axial-vector form factors in Eqs.~\eqref{eq.Vmunu} and \eqref{eq.Amunu}. The explicit expressions of form factors $v_1,v_2,v_3,v_4$ are given as follows:  
\begin{eqnarray}
v_1^{\rm VVP}&=&2\sqrt2 g_{\rm{\omega}} D_\omega[(k+p_2)^2] \bigg\{
 \frac{-2\sqrt2 F_V }{ F    } D_\rho[(k+p_1+p_2)^2]  
[ 4d_3 k \cdot p_2 +(d_1+8d_2+2d_3)m_\pi^2 \nonumber \\ 
&&+2d_3 (p_1\cdot p_2  + k\cdot p_1)] + g_\omega  -\frac{2 F_V }{ F } D_\rho[k^2]
[ 2d_3 k \cdot p_2+ (d_1+8d_2)m_\pi^2 ] \bigg\},  \\
v_2^{\rm VVP}&=&-\frac{2\sqrt2 g_{\rm{\omega}}}{M_\omega^2}  D_\omega[(k+p_2)^2]\bigg\{
 \frac{2\sqrt2 F_V }{ F } D_\rho[(k+p_1+p_2)^2] 
  [ (d_4-d_3) M_\omega^2 (k\cdot p_1+ p_1\cdot p_2) \nonumber \\ 
 && -(d_1+8d_2+d_3+d_4) m_\pi^2 ( p_1\cdot p_2 + k\cdot p_1)  -2(d_3+d_4) k\cdot p_2 ( p_1\cdot p_2+ k\cdot p_1 ) 
   \nonumber \\
  &&-(d_1+8d_2-d_3+d_4) m_\pi^2 ( m_\pi^2-M_\omega^2 +2 k\cdot p_2 ) -2d_3 (p_1\cdot p_2+k\cdot p_1)^2 ] \nonumber\\
&&+ g_{\rm{\omega}}(m_\pi^2-M_\omega^2+2k\cdot p_2 + p_1\cdot p_2 + k\cdot p_1 ) \nonumber \\
&&+ \frac{2 F_V }{ F } D_\rho[k^2]
\{-(m_\pi^2-M_\omega^2+2 k\cdot p_2)[ -2d_4 k \cdot p_2+ (d_1+8d_2-d_3-d_4)m_\pi^2 ]
\nonumber \\
&&-(k\cdot p_1+ p_1\cdot p_2)[ (d_1+8d_2-d_3-d_4)m_\pi^2+(d_3+d_4)M_\omega^2-2d_4 k \cdot p_2 ]\}  \bigg\},    \\
 v_3^{\rm VVP}&=&\frac{2\sqrt2 g_{\rm{\omega}}}{M_\omega^2 } D_\omega[(k+p_2)^2]\bigg\{ \frac{-2\sqrt2 F_V}{ F } D_\rho[(k+p_1+p_2)^2]   [  ( d_4-d_1-8d_2-d_3)m_\pi^2 \nonumber \\ 
 &&  - (d_3+d_4 ) M_\omega^2  + 2(d_4-d_3)  k\cdot p_2   - 2 d_3 ( k\cdot p_1+ p_1\cdot p_2)  ]\nonumber \\
&&-g_\omega  +\frac{2 F_V }{ F  }\, D_\rho[k^2][-2d_4 k \cdot p_2+ (d_1+8d_2-d_3-d_4)m_\pi^2 + (d_3+d_4)M_\omega^2 ] \bigg\}\,, \nonumber \\
 v_4^{\rm VVP}&=&\frac{2\sqrt2 g_{\rm{\omega}}}{M_\omega^2 } D_\omega[(k+p_2)^2] \bigg\{ g_\omega
 -\frac{2 F_V }{ F } D_\rho[k^2] [ (d_1+8d_2)m_\pi^2 + (d_3+d_4)(M_\omega^2 -m_\pi^2)
  \nonumber\\
 &&-2d_4 k \cdot p_2]-\frac{4 F_V }{ \sqrt2F } D_\rho[(k+p_1+p_2)^2] 
 [( d_1+8d_2)m_\pi^2 + 2 d_3  (M_\omega^2+k\cdot p_1+ p_1\cdot p_2 ) ] \bigg\}\,,\nonumber\\
\end{eqnarray}
where the quantity of $D_R$ for the resonance $R$ is defined as
\begin{eqnarray}\label{eq.defprogd}
D_R(s)= \frac{1}{M_R^2-s-i M_R \Gamma_R(s)}\,.
\end{eqnarray}
For the broad $\rho$ resonance, we use the energy dependent decay width~\cite{Guerrero:1997ku}
\begin{eqnarray}\label{eq.widthrho}
\Gamma_\rho(s)= \frac{M_\rho s}{96\pi F_\pi^2} \bigg[ \bigg( 1- \frac{4m_\pi^2}{s} \bigg)^{\frac{3}{2}}\theta(s-4m_\pi^2) +\frac{1}{2}\bigg( 1- \frac{4m_K^2}{s} \bigg)^{\frac{3}{2}} \theta(s-4m_K^2) \bigg]\,, 
\end{eqnarray}
with $\theta(x)$ the Heaviside function. For the narrow $\omega$, we simply take the constant width in Eq.~\eqref{eq.defprogd}. 
For the pion, it is simply given by
\begin{eqnarray}
D_\pi(s)= \frac{1}{m_\pi^2-s}\,.
\end{eqnarray}

For the axial-vector form factors, we can obtain the corresponding expressions by evaluating the Feynman diagrams in Fig.~\ref{fig.a} and the results read 
\begin{eqnarray}
a_1^{\rm VVP}&=&\frac{2 g_{\rho}}{F M_V^2}\bigg\{
2 G_V M_V^2 k\cdot p_2 D_\pi[(k+p_1+p_2)^2] \{ D_\rho[(k+p_1)^2] +D_\rho[(k+p_2)^2] \}\nonumber \\
&&+D_\rho[(k+p_1)^2]\big[F_V M_V^2 +  (F_V-2G_V) p_1\cdot p_2 \big]\nonumber\\
&&-D_\rho[(k+p_2)^2]\,(F_V- 2G_V)(M_V^2-m_\pi^2-k\cdot p_2) \bigg\} ,\\
a_2^{\rm VVP}&=&\frac{2 g_{\rho}}{F M_V^2}\bigg\{
-2 G_V M_V^2 k\cdot p_1 D_\pi[(k+p_1+p_2)^2] \{ D_\rho[(k+p_1)^2] +D_\rho[(k+p_2)^2] \}\nonumber \\
&&+D_\rho[(k+p_1)^2](F_V- 2G_V)(M_V^2-m_\pi^2-k\cdot p_2) \nonumber \\
&&-D_\rho[(k+p_2)^2]\big[F_V M_V^2 +  (F_V-2G_V) p_1\cdot p_2 \big] \bigg\} ,\\
a_3^{\rm VVP}&=&\frac{2 g_{\rho}}{F M_V^2}\bigg\{
-2G_V M_V^2 D_\pi[(k+p_1+p_2)^2]\{D_\rho[(k+p_1)^2]+D_\rho[(k+p_2)^2]\}\nonumber \\
&&+D_\rho[(k+p_1)^2](F_V- 2G_V) \bigg\} ,\nonumber \\
a_4^{\rm VVP}&=&\frac{2 g_{\rho}}{F M_V^2}\bigg\{
-2G_V M_V^2 D_\pi[(k+p_1+p_2)^2]\{D_\rho[(k+p_1)^2]+D_\rho[(k+p_2)^2]\}\nonumber \\
&&+D_\rho[(k+p_2)^2](F_V- 2G_V) \bigg\} \,.
\end{eqnarray}

The form factors $a_1, a_2$ in Ref.~\cite{Cirigliano:2002pv} can be also given in terms of the new bases introduced in Eq.~\eqref{eq.Amunu} 
\begin{eqnarray}
a_1^{\rm CEN}&=&-\frac{1}{8\pi^2 F^2}-\frac{1}{4\pi^2 F^2}\,D_\pi[(k+p_1+p_2)^2] \, k\cdot p_2 \,,\nonumber \\
a_2^{\rm CEN}&=&\frac{1}{8\pi^2 F^2}+\frac{1}{4\pi^2 F^2}\,D_\pi[(k+p_1+p_2)^2] \, k\cdot p_1  \,,\nonumber \\
a_3^{\rm CEN}&=&\frac{1}{4\pi^2 F^2}\,D_\pi[(k+p_1+p_2)^2]  \,,\nonumber \\
a_4^{\rm CEN}&=&\frac{1}{4\pi^2 F^2}\,D_\pi[(k+p_1+p_2)^2]  \,.
\end{eqnarray}
Finally, we will use $v_i^{\rm Our}=v_i^{\rm CEN}+v_i^{\rm VVP} ,a_i^{\rm Our}=a_i^{\rm CEN}+a_i^{\rm VVP}$ to calculate the various quantities  discussed in the Sec.~\ref{sec.pheno}. The form factors contributed by the minimal resonance chiral Lagrangians, denoted by $v_i^{\rm CEN}$ and $a_i^{\rm CEN}$, have been explicitly given in Ref.~\cite{Cirigliano:2002pv} and we do not show them again here.

After taking into account the constraints of the high-energy behaviors and the on-shell Feynman rules, there is only one undetermined parameter $d_4$, which can be fixed via the $\omega\to\pi\pi\gamma$ decay width. With the $VVP$ and $VJP$ Lagrangians in Eqs.~\eqref{eq.lagvjp} and \eqref{eq.lagvvp}, the only relevant Feynman diagram to the $\omega \rightarrow \pi^0\pi^0\gamma$ decay amplitude corresponds to the $\rho^0$ mediated one. Under the on-shell approximation of the $JV\pi$ transition vertex, the decay amplitude of the $\omega(q) \to \pi^0(p_1)\pi^0(p_2)\gamma(k)$ process reads 
\begin{eqnarray}
T_{\omega\rightarrow \pi^0\pi^0\gamma}=&& \frac{2}{F} \bigg\{\,\, d_1(\epsilon_{\lambda\delta\mu\sigma} p_{1\nu} p_1^{\sigma}
+ \epsilon_{\mu\nu\lambda\sigma} p_{1\delta} p_1^{\sigma} ) + 4d_2 m_\pi^2\epsilon_{\mu\nu\lambda\delta}
\nonumber \\&&
+ d_3 [\,\epsilon_{\lambda\delta\mu\sigma} (k+p_2)_{\nu} p_1^{\sigma}- \epsilon_{\mu\nu\lambda\sigma} q_{\delta} p_1^{\sigma} \, ]
+d_4 [\,\epsilon_{\lambda\delta\mu\sigma} (k+p_2)^{\sigma} p_1^{\nu}- \epsilon_{\mu\nu\lambda\sigma} q^{\sigma} p_{1\delta} ]
\,\bigg\} \nonumber \\ && D^{\lambda\delta,\beta\theta}(k+p_2, M_V^2) \,e\, \frac{g_{\omega}}{3}\,\epsilon_{\beta\theta\rho\alpha}k^\rho
\varepsilon^\alpha_\gamma(k) \, \, \frac{q^\mu \varepsilon^\nu_\omega(q)- q^\nu \varepsilon^\mu_\omega(q)}{M_\omega} 
+ \bigg(p_1 \leftrightarrow p_2 \bigg)\,,\nonumber \\
\end{eqnarray}
where $\varepsilon^\mu_\omega(q)$ stands for the polarization vector of the $\omega$ resonance, the propagator with the anti-symmetric tensor is defined as 
\begin{eqnarray}
D^{\mu \nu, \rho \sigma}(k, M_{V}^{2})=\frac{1}{M_{V}^{2} D_{\rho}(k^{2})}\bigg[g^{\mu \rho} g^{\nu \sigma}\left(M_{V}^{2}-k^{2}\right)+g^{\mu \rho} k^{\nu} k^{\sigma}-g^{\mu \sigma} k^{\nu} k^{\rho}-\big(\mu \longleftrightarrow \nu\big)\bigg]\,, 
\end{eqnarray}
and the propagator $D_\rho(t)$ with the energy dependent width is given in  Eq.~\eqref{eq.widthrho}. 

It is then straightforward to calculate the $\omega\to\pi^0\pi^0\gamma$ decay width
\begin{eqnarray}
\Gamma_{\omega\rightarrow \pi^0\pi^0\gamma}=\int_{4m_\pi^2}^{M_\omega^2} d s \int_{t_{-}}^{t_{+}} d t
\frac{1}{(2\pi)^3}\frac{1}{32M_\omega^3} \frac{1}{3} |T_{\omega\rightarrow \pi^0\pi^0\gamma}|^2\,,
\end{eqnarray}
where the standard Mandelstam variables are defined as 
\begin{eqnarray}
 s= (p_1+p_2)^2\,, \qquad t=(k+p_2)^2\,,
\end{eqnarray}
and the upper and lower limits of the integrals are  
\begin{eqnarray}
t_{\pm}&=& \frac{M_\omega^2+2m_\pi^2-s}{2}\pm \frac{\sqrt{s^2-4m_\pi^2s}(M_\omega^2-s)}{2s}\,. 
\end{eqnarray}

By using the PDG result of the decay width~\cite{Zyla:2020zbs} 
\begin{equation}
\Gamma_{\omega\rightarrow \pi^0\pi^0\gamma}^{Exp}=(5.8\pm 1.0) \times 10^{-4}~{\rm MeV}\,,
\end{equation}
we can determine the values of $d_4$ and two distinct solutions are found 
\begin{eqnarray}\label{eq.d4value}
d_4&=&-0.12 \pm 0.05\,, \nonumber \\
d_4&=&0.82 \pm 0.05\,,
\end{eqnarray}
by taking into account the high-energy constraints in Eq.~\eqref{eq.fvgv2}. \newtex{It is noted that a guesstimate range of $|d_4|<0.15$ is advocated in Ref.~\cite{Miranda:2020wdg}. This ballpark estimation of $d_4$ gives preference of the negative value over the positive one in Eq.~\eqref{eq.d4value}. Moreover, we could also determine the values of $d_4$ by using another set of high-energy constraints~\eqref{eq.fvgv3}, and this can give another two solutions to $d_4$: the negative one around $-0.4$ and the positive one around $0.8$, both of which do not fall in the ballpark estimate range of Ref.~\cite{Miranda:2020wdg}. Therefore we refrain from discussing the latter two solutions of $d_4$. We will then focus on the two solutions in Eq.~\eqref{eq.d4value} and study their impacts on the $\tau\to\pi\pi\gamma\nu_\tau$ in the phenomenological discussions. Our result shows that many curves obtained with $d_4=0.82$, such as those in Figs.~\ref{fig.pipi}, \ref{fig.pig} and \ref{fig.xiall}, are significantly different from the corresponding ones by taking the minimal resonance chiral operators in Ref.~\cite{Cirigliano:2002pv}. This indicates that the value of $d_4$ with larger magnitude in Eq.~\eqref{eq.d4value} does not likely correspond to a physically meaningful solution, since it is not expected that the $VVP$ effects are so huge with respect to the leading contributions in resonance chiral theory. }

\section{Kinematical factors in the invariant amplitudes squared}\label{sec.kinf}
\setcounter{equation}{0}

In Eqs.~\eqref{eq.m02} and \eqref{eq.m12}, we have introduced several types of  kinematical factors in different parts of the amplitude squared $|\mathcal{M}|^2$. During the calculation, the FeynCalc package~\cite{feyncalc} is used to crosscheck the following formulas. Here we give the explicit expressions of the various factors for completeness. 

First, the relevant Lorentz scalars constructed by the momenta of the particles in the $\tau^-(P)\to\pim(p_1)\pin(p_2)\nu_\tau(q)\gamma(k)$ process are defined as follows 
\begin{eqnarray}\label{eq.defkinv}
	P^2=m_0^2,\, q^2=0,\, k^2=0,\quad
	P\cdot p_1=d m_0^2,\quad   P\cdot p_2=c m_0^2, \nonumber\\
	P\cdot q=f m_0^2,\quad P\cdot k=g m_0^2,\quad p_1\cdot p_2=h m_0^2,\quad p_1\cdot q=j m_0^2,\nonumber\\
	p_1\cdot k=n m_0^2,\quad p_2\cdot q=l m_0^2,\quad p_2\cdot k=m m_0^2,\quad q\cdot k=w m_0^2 \,.
\end{eqnarray}

The effect of isospin violation is ignored in this paper, so we have $p_1^2=r_1 m_0^2= p_2^2=r_2 m_0^2=r m_0^2$. The kinematical factors introduced in Eqs.~\eqref{eq.m02} and \eqref{eq.m12} are 

\begin{eqnarray}
	C_{f_t f_t}&=&
	(2(4d^2g(j-l)(n+m)+d(-2g^2(j-l)(2h+r_1+r_2)\nonumber\\
	&&-g(n+m)(w(2h-r_1-r_2)+4jn-4nl)-2(j-l)(n+m)^2)\nonumber\\
	&&-4c^2g(j-l)(n+m)+c(2g^2(j-l)(2h+r_1+r_2)\nonumber\\
	&&+g(n+m)(w(r_1+r_2-2h)+4(j-l)m)+2(j-l)(n+m)^2)\nonumber\\
	&&+(n+m)(g^2(2h(j+l)+jr_1-3jr_2-3lr_1+lr_2)\nonumber\\
	&&-g(n+m)(-2hw-2jn+2jm+2nl-2lm+r_1w+r_2w)\nonumber\\
	&&-(n+m)(-2hw-2jn+2jm+2nl-2lm+r_1w+r_2w)))\nonumber\\
	&&-f(2h-r_1-r_2)(-2g(n+m)(2d+2c-n-m)\nonumber\\
	&&+2g^2(2h+r_1+r_2)+2(n+m)^2))/(4g^2(n+m)^2)\,,
\end{eqnarray}

\begin{eqnarray}
	C_{f_u f_u}&=&
	(2d(2hjnm+hn^2w-2hnlm+hnmw+2jnnm-jn^2r_2+2jnm^2-jm^2r_1-n^3l\nonumber\\
	&&+n^3w-2n^2lm+n^2lr_2+2n^2mw-nlm^2+nm^2w-nmr_1w+lm^2r_1-m^2r_1w)\nonumber\\
	&&-2c(j(nm(2h+m)+n^3+n^2(2m-r_2)-m^2r_1)+n^2(hw-2lm+lr_2)\nonumber\\
	&&+nm(-2hl+hw-r_1w)-2n^3l+m^2r_1(l-w))+4fh^2nm+4fhn^2m\nonumber\\
	&&-2fhn^2r_2+4fhnm^2-2fhnmr_1-2fhnmr_2-2fhm^2r_1-2fn^3r_2\nonumber\\
	&&-2fn^2mr_2+fn^2r_1r_2+fn^2r_2^2-2fnm^2r_1-2fm^3r_1+fm^2r_1^2+fm^2r_1r_2\nonumber\\
	&&+2g(n+m)(hn(j-l)+j(n^2+nm-mr_1)-r_1w(n+m)+lmr_1))\nonumber\\
	&&/(2n^2(n+m)^2)\,,
\end{eqnarray}

\begin{eqnarray}
	C_{f_t f_u}&=&-
	(n^2(-2cjn+4gjn+2jn+2cln-2gln-2ln-4c^2j+2ghj+4c^2l+4cgl\nonumber\\
	&&+6gjm+4jm-4glm-4lm-glr_1-2gjr_2+glr_2-2chw+2ghw+cr_1w\nonumber\\
	&&-2gr_1w+cr_2w)+n(2cjmm+2jm^2-2clm^2-2glm^2-2lm^2+2g^2hj\nonumber\\
	&&+4cghj-2g^2hl-4cghl-4c^2jm+2ghjm+4c^2lm+4cglm-2ghlm\nonumber\\
	&&+2g^2jr_1-2g^2lr_1-gjmr_1+glmr_1+4cgjr_2-4cglr_2-gjmr_2+glmr_2\nonumber\\
	&&-2cghw-2chmw+4ghmw-2cgr_1w+cmr_1w-4gmr_1w+cmr_2w)\nonumber\\
	&&-2gjm^3-2ghlm^2+2g^2hjm-4cghjm-2g^2hlm+4cghlm-gjm^2r_1\nonumber\\
	&&+2glmmr_1+2g^2jmr_1-4cgjmr_1-2g^2lmr_1+4cglmr_1+gjm^2r_2\nonumber\\
	&&-2fg(2(n-m)h^2+(n+m)(n-m-r_1+r_2)h+n^2r_2\nonumber\\
	&&+mr_1(-m+r_1+r_2)-n(m(r_1-r_2)+r_2(r_1+r_2)))\nonumber\\
	&&+f(n+m)(2cn(2h-r_1-r_2)+2d(n^2+m(-2h-m+r_1+r_2)))\nonumber\\
	&&+2gmw(hm-ch-mr_1-cr_1)-2d^2(n+m)(2jm-2lm+(n+m)w)\nonumber\\
	&&+d(n+m)(2lm^2-2nlm+2j(n-m)m+2hwm-r_1wm-r_2wm\nonumber\\
	&&+2c(n+m)(2j-2l+w))\nonumber+2dg(ln^2+2lmn+2lr_2n+r_1wn\nonumber\\
	&&+lm^2-2lmr_1-j(3n^2+4mn+2r_2n+mm-2mr_1)+mr_1w\nonumber\\
	&&+h(-2jn+2ln+wn+2jm-2lm+mw)))/(2gn(n+m)^2)\,,
\end{eqnarray}

\begin{eqnarray}
	\tilde{C}_{f_t f_u}&=&
	(n(4d-3g+2h-2j+n+2l+2m-r_1-r_2+w)-4g(h+r_1)\nonumber\\
	&&+m(4d+g+2h+2j-2l+m-r_1-r_2+w))/(2gn(n+m)) \,,
\end{eqnarray}

\begin{eqnarray}
	C_{v_1 v_1}&=&(2dnw+2gjn-2gr_1w)m_0^4/2 \,,\nonumber\\
	C_{v_2 v_2}&=&(2cmw+2glm-2gr_2w)m_0^4/2 \,,\nonumber\\
	C_{v_3 v_3}&=&-(2dj-fr_1)(-2hnm+n^2r_2+m^2r_1)m_0^8/2 \,,\nonumber\\
	C_{v_4 v_4}&=&-m_0^8/2(2hnm+n^2(-r_2)-m^2r_1)  \nonumber\\
	&&(f(2h+2n+2m+r_1+r_2)-2(d+c+g)(j+l+w)) \,,\nonumber \\
	C_{v_1 v_2}&=&m_0^4(dmw+cnw+g(-2hw+jm+nl)) \,,\nonumber\\
	C_{v_1 v_3}&=&m_0^6(d(-hnw-2jnm+n^2l+mr_1w)+cjn^2+g(jmr_1-hjn))\,,\nonumber\\
	C_{v_1 v_4}&=&-m_0^6(dhnw+2djnm-dn^2l+dnlm+dnmw-dmr_1w -2fhnm\nonumber\\
	&&-c(-hnw+jn(n-m)+n^2(2l+w)+mr_1w)+fn^2r_2+fm^2r_1\nonumber\\
	&&+g(hn(j+l+2w)+jnm-mr_1(j+l+2w)+n^2(-l)))\,,\nonumber\\
	C_{v_2 v_3}&=&m_0^6(dhmw-2djm^2+dnlm-dnr_2w+cjnm\nonumber\\
	&&+f(-2hnm+n^2r_2+m^2r_1)+gj(hm-nr_2))\,,\nonumber\\
	C_{v_2 v_4}&=&m_0^6(d(hmw-2jm^2+nlm-nr_2w-lm^2-m^2w)+chmw+cjnm\nonumber\\
	&&-cjm^2+2cnlm+cnmw-cnr_2w+f(-2hmn+nr_2 n+m^2r_1)\nonumber\\
	&&+g(hm(j+l+2w)-nr_2(j+2w)-jm^2+nl(m-r_2)))\,,\nonumber\\
	C_{v_3 v_4}&=&-m_0^8(2hnm+n^2(-r_2)-m^2r_1)\nonumber\\
	&&(-d(2j+l+w)+j(-(c+g))+f(h+n+r_1))\,,
\end{eqnarray}

\begin{eqnarray}
	\tilde{C}_{v_1 v_2}&=&-(n+m)m_0^4\,,\nonumber\\
	\tilde{C}_{v_2 v_3}&=&-(hm-n(m+r_2))m_0^6\,,\nonumber\\
	\tilde{C}_{v_1 v_3}&=&(hn+n^2-mr_1)m_0^6\,,
\end{eqnarray}

\begin{equation}
	C_{f_a I_i}=C_{f_a I_i}^{F}+C_{f_a I_i}^{SI},\quad
	\tilde{C}_{f_a I_i}=\tilde{C}_{f_a I_i}^{F}+\tilde{C}_{f_a I_i}^{SI}
\end{equation}
with $a=t,u$ and $I_{i=1,2,3,4}=v_{i=1,2,3,4}, a_{i=1,2,3,4}$.
The explicit expressions for these factors are 
\begin{eqnarray}
	\tilde{C}_{f_t v_1}^{F}&=&-(4d+g+n+m+w)m_0^2/(2g)\,,\nonumber\\
	\tilde{C}_{f_t v_2}^{F}&=&-(4c+g+n+m+w)m_0^2/(2g)\,,\nonumber\\
	\tilde{C}_{f_t v_3}^{F}&=&(-dm+cn+h(n+m)-2jn+jm+nl-nr_1-mr_1)m_0^4/(2g)\,,\nonumber\\
	\tilde{C}_{f_t v_4}^{F}&=&((2d-2c+g)(m-n)+n(-2j-n-r_1+r_2-w)\nonumber\\
	&&+m(m-r_1+r_2+w+2l))m_0^4/(2g)\,,
\end{eqnarray}

\begin{eqnarray}
	C_{f_t v_1}^{F}&=&
	-(2d^2w+d(-2cw-2fn+2fm+2g(j-l))\nonumber\\
	&&-j(4gn-2gm+2n)-2ghw+2gnl+2gr_1w+2nl)m_0^2/(2g)\,,\nonumber\\
	C_{f_t v_2}^{F}&=&
	(c(-2dw+2fn-2fm-2gj+2gl)\nonumber\\
	&&+2c^2w+2g(-hw+jm+nl-2lm+r_2w)+2jm-2lm)m_0^2/(2g)\,,\nonumber\\
	C_{f_t v_3}^{F}&=&
	(4d^2 jm-2d^2 lm+c(-2dj(2n+m)+2dnl+2fn(r_1-h)\nonumber\\
	&&+(n-m)(2jn-r_1w))+2dfhm-2dfmr_1-2dhmw-2djnm\nonumber\\
	&&+2djm^2+dnr_2w+dmr_1w+2c^2jn+2fhnm-fn^2r_2-fm^2r_1\nonumber\\
	&&+g(-2hjn+n(jr_2+lr_1)+mr_1(j-l)))m_0^4/(2g)\,,\nonumber\\
	C_{f_t v_4}^{F}&=&
	(2d^2m(2j+w)-d(2c(2jn+w(n+m)+2lm)+2fnm-2fm^2\nonumber\\
	&&+2fmr_1-2fmr_2+2gm(l-j)+2hmw+2jnm-2jm^2+2nlm\nonumber\\
	&&-2nr_2w-2lm^2-mr_1w+mr_2w)+2c^2n(2l+w)\nonumber\\
	&&+c(2fn(n-m+r_1-r_2)+2gn(l-j)-2hnw+2jn^2-2jnm\nonumber\\
	&&+2n^2l-2nlm-nr_1w+nr_2w+2mr_1w)+4fhnm-2fn^2r_2-2fm^2r_1\nonumber\\
	&&-2ghjn-2ghnw-2ghlm-2ghmw-2gjnm+2gjm^2+gjmr_1+gjmr_2\nonumber\\
	&&+2gn^2l-2gnlm+gnlr_1+gnlr_2+2gnr_2w+2gmr_1w)m_0^4/(2g)\,,
\end{eqnarray}

\begin{eqnarray}
	\tilde{C}_{f_t v_1}^{SI}&=&2(h+n+r_1)m_0^2/(n+m)\,,\nonumber\\
	\tilde{C}_{f_t v_2}^{SI}&=&2(h+m+r_2)m_0^2/(n+m)\,,\nonumber\\
	\tilde{C}_{f_t v_3}^{SI}&=&-(h(n-m)+nr_2-mr_1)m_0^4/(n+m)\,,\nonumber\\
	\tilde{C}_{f_u v_1}^{SI}&=&-(2hn+n^2-nm-2mr_1)m_0^2/(n^2+nm)\,,\nonumber\\
	\tilde{C}_{f_u v_2}^{SI}&=&(2m(h+m)+n^2+n(m-2r_2))m_0^2/(n(n+m))\,,\nonumber\\
	\tilde{C}_{f_u v_3}^{SI}&=&-(hn(n+3m)+n^3+n^2(m-r_2)-nmr_1-2m^2r_1)m_0^4/(n(n+m))\,,\nonumber\\
\end{eqnarray}

\begin{eqnarray}
	C_{f_t v_1}^{SI}&=&
	(dhw-2djn+dr_1w-chw+2cnl-cr_1w\nonumber\\
	&&+f(-hn+hm-nr_2+mr_1)+g(h+r_1)(j-l))m_0^2/(n+m)\,,\nonumber\\
	C_{f_t v_2}^{SI}&=&(dhw-2djm+dr_2w-chw+2clm-cr_2w\nonumber\\
	&&+f(-hn+hm-nr_2+mr_1)+g(h+r_2)(j-l))m_0^2/(n+m)\,,\nonumber\\
	C_{f_t v_3}^{SI}&=&(h(n-m)+nr_2-mr_1)(2dj-dl-cj+f(h-r_1))m_0^4/(n+m)\,,\nonumber\\
	C_{f_t v_4}^{SI}&=&(d(2j+w)-2cl-cw-fn+fm-fr_1+fr_2+g(j-l))\nonumber\\
	&&(h(n-m)+nr_2-mr_1)m_0^4/(n+m)\,,
\end{eqnarray}

\begin{eqnarray}
	C_{f_u v_1}^{SI}&=&
	(d(-hnw-2jnm+n^2l-2n^2w+nlm-2nmw+mr_1w)-2fhnm\nonumber\\
	&&+c(hnw+jn(n+m)-2n^2l-mr_1w)+fn^2r_2+fm^2r_1+g(hn(l-j)\nonumber\\
	&&+j(-2n^2-2nm+mr_1)+2r_1w(n+m)-lmr_1))m_0^2/(n(n+m))\,,\nonumber\\
	C_{f_u v_2}^{SI}&=&-
	(-dhmw+2djm^2-dnlm+dnmw+dnr_2w-dlm^2+dm^2w\nonumber\\
	&&+chmw-cjnm-cjm^2+cn^2w+2cnlm+cnmw-cnr_2w\nonumber\\
	&&+f(2hnm+n^2(-r_2)-m^2r_1)+g(n(-2hw+j(m+r_2)+lm-lr_2)\nonumber\\
	&&+m(h(-j+l-2w)+jm)+n^2l))m_0^2/(n(n+m))\,,\nonumber\\
	C_{f_u v_3}^{SI}&=&
	(4dhjnm+dhn^2w-2dhnlm+dhnmw+2djn^2m-2djn^2r_2\nonumber\\
	&&+2djnm^2-2djm^2r_1-dn^3l-dn^2lm+dn^2lr_2-dnmr_1w\nonumber\\
	&&+dlm^2r_1-dm^2r_1w-cj(2hnm+n^3+n^2(m-r_2)-m^2r_1)\nonumber\\
	&&+2fh^2nm-fhn^2r_2-2fhnmr_1-fhm^2r_1+fn^2r_1r_2\nonumber\\
	&&+fm^2r_1^2+gj(n+m)(hn-mr_1))m_0^4/(n(n+m))\,,\nonumber\\
	C_{f_u v_4}^{SI}&=&
	(4dhjnm+dhn^2w+3dhnmw+2djn^2m-2djn^2r_2+2djnm^2\nonumber\\
	&&-2djm^2r_1-dn^3l+dn^2mw-dn^2r_2w+dnlm^2+dnm^2w\nonumber\\
	&&-dnmr_1w-2dm^2r_1w-c(n^2(2l(m-r_2)-w(h-m+r_2))\nonumber\\
	&&+nm(h(4l+w)+r_1w)+j(n^3-nm^2)+n^3(2l+w)-2lm^2r_1)\nonumber\\
	&&-4fhn^2m-2fhnmr_1+2fhnmr_2+2fn^3r_2+fn^2r_1r_2-fn^2r_2^2\nonumber\\
	&&+2fnm^2r_1+fm^2r_1^2-fm^2r_1r_2+g(hn(j(n+3m)+nl+2nw\nonumber\\
	&&-lm+2mw)+n^2(j-l)(m-r_2)+nm(j(m-r_1)-r_1(l+2w))\nonumber\\
	&&-2m^2r_1(j+w)+n^3(-l)))m_0^4/(n(n+m))\,,
\end{eqnarray}

\begin{eqnarray}
	\tilde{C}_{f_t a_1}^{F}&=&-(d-c-f-g)m_0^2/g\,,\nonumber\\
	\tilde{C}_{f_t a_2}^{F}&=&(d-c+f+g)m_0^2/g\,,\nonumber\\
	\tilde{C}_{f_t a_3}^{F}&=&-m_0^4/(2g)(4dj-2dl+dm-2cj-cn+2f(h-r_1)\nonumber\\
	&&+g(h-r_1)-hw-2jn+jm+nl+r_1w)\,,\nonumber\\
	\tilde{C}_{f_t a_4}^{F}&=&m_0^4/(2g)(-2dl-dm-2cj+cn+4cl+2f(h-r_2)\nonumber\\
	&&+g(h-r_2)-hw+jm+nl-2lm+r_2w)\,,
\end{eqnarray}

\begin{eqnarray}
	C_{f_t a_1}^{F}&=&
	(-wd^2+f(n-m)d+(gj+cw)d-cgj-2gjn-jn\nonumber\\
	&&+gnl+gjm+jm+fg(h-r_1)+w(-gh-h+gr_1+r_1))m_0^2/g\,,\nonumber\\
	C_{f_t a_2}^{F}&=&
	-(-wc^2+glc+f(m-n)c+dwc-dgl+gnl+nl\nonumber\\
	&&+gjm-2glm-lm+fg(h-r_2)-ghw-hw+gr_2w+r_2w)m_0^2/g\,,\nonumber\\
	C_{f_t a_3}^{F}&=&
	-(2lmd^2-2r_2wd^2+2jm^2d-2ghld-2jnmd+2gjr_2d-2hmwd\nonumber\\
	&&+mr_1wd+nr_2wd-2ghjn+2hnl+2hjm+gnlr_1+gjmr_1-glmr_1\nonumber\\
	&&-2lmr_1+gjnr_2-2jnr_2-f(-2gh^2+2cnh+2dmh-2nmh\nonumber\\
	&&+m^2r_1-2cmr_1+n^2r_2-2dnr_2+2gr_1r_2)-2h^2w+2r_1r_2w\nonumber\\
	&&+2c^2(jn-r_1w)+c(-2dnl+g(2lr_1-2hj)+w(4dh-nr_1+mr_1)\nonumber\\
	&&+j(2n^2-2mn-2dm)))m_0^4/(2g)\,,\nonumber\\
	C_{f_t a_4}^{F}&=&
	-(2lmd^2-2r_2wd^2+2lm^2d-2ghld-2nlmd+2gjr_2d+nr_2wd\nonumber\\
	&&-mr_2wd+2hnl+2hjm-2ghlm+glmr_1-2lmr_1-gjnr_2\nonumber\\
	&&-2jnr_2+gnlr_2+gjmr_2-f(-2ghh+2cnh+2dmh-2nmh\nonumber\\
	&&+m^2r_1-2cmr_1+n^2r_2-2dnr_2+2gr_1r_2)-2h^2w+2r_1r_2w\nonumber\\
	&&+2c^2(jn-r_1w)+c(2ln^2-2dln-2lmn-2hwn+r_2wn\nonumber\\
	&&-2ghj-2djm+2glr_1+4dhw+mr_1w))m_0^4/(2g)\,,
\end{eqnarray}

\begin{eqnarray}
	\tilde{C}_{f_t a_1}^{SI}&=&
	-\tilde{C}_{f_t a_2}^{SI}=
	-\tilde{C}_{f_u a_1}^{SI}
	=(d-c+j-l)m_0^2/(n+m)\,,\nonumber\\
	\tilde{C}_{f_u a_2}^{SI}&=&-(dm-cm+g(n+m)+jm+nw-lm+mw)m_0^2/(n(n+m))\,,\nonumber\\
	\tilde{C}_{f_u a_3}^{SI}&=&-(d+j)m_0^4\,,\nonumber\\
	\tilde{C}_{f_u a_4}^{SI}&=&-(c+l)m_0^4\,,
\end{eqnarray}

\begin{eqnarray}
	C_{f_t a_1}^{SI}&=&-C_{f_t a_2}^{SI}=
	(d(hw-nl+lm-r_2w)+c(hw+jn-jm-r_1w)\nonumber\\
	&&\qquad \qquad \quad+g(-h(j+l)+jr_2+lr_1))m_0^2/(n+m)\,,\nonumber\\
	C_{f_u a_1}^{SI}&=&
	(dw(-h-2m+r_2)+dn(l-2w)-dlm+cw(r_1-h)\nonumber\\
	&&+cj(m-n)+gh(j+l)+gj(2n+2m-r_2)-glr_1)m_0^2/(n+m)\,,\nonumber\\
	C_{f_u a_2}^{SI}&=&
	m_0^2(d(n(lm-mw)-m(w(h+m-r_2)+lm))\nonumber\\
	&&-c(w(m(h-r_1)+n^2+nm)+jm(n-m))\nonumber\\
	&&+g(m(h(j+l)+jm-jr_2-lr_1)+n(mj+ml+nl)))/(n^2+nm)\,,\nonumber\\
	C_{f_u a_3}^{SI}&=&m_0^4(dhw-dnl+cjn-cr_1w-ghj+glr_1)\,,\nonumber\\
	C_{f_u a_4}^{SI}&=&m_0^4(-dlm+dr_2w-chw+cjm+ghl-gjr_2)\,,
\end{eqnarray}

\begin{eqnarray}
	\tilde{C}_{v_1 a_2}&=&-\tilde{C}_{v_2 a_1}=m_0^4(g+w)\,,\nonumber\\
	\tilde{C}_{v_1 a_3}&=&-\tilde{C}_{v_3 a_1}=m_0^6 n(d+j)\,,\nonumber\\
	\tilde{C}_{v_2 a_3}&=&-\tilde{C}_{v_3 a_2}=m_0^6 m(d+j)\,,\nonumber\\
	\tilde{C}_{v_1 a_4}&=&m_0^6 n(c+l)\,,\nonumber\\
	\tilde{C}_{v_2 a_4}&=&m_0^6 m(c+l)\,,\nonumber\\
	\tilde{C}_{v_4 a_1}&=&-n(d+c+g+j+l+w)m_0^6\,,\nonumber\\
	\tilde{C}_{v_4 a_2}&=&m(d+c+g+j+l+w)m_0^6\,,
\end{eqnarray}

\begin{eqnarray}
	C_{v_1 a_1} &=&-2n m_0^4(gj-dw)\,,\nonumber\\
	C_{v_2 a_2} &=&-2m m_0^4(gl-cw)\,,\nonumber\\
	C_{v_1 a_2} &=&-m_0^4(-dmw-cnw+gjm+gnl)\,,\nonumber\\
	C_{v_2 a_1} &=&-m_0^4(-dmw-cnw+gjm+gnl)\,,\nonumber\\
	C_{v_1 a_3} &=&C_{v_3 a_1}=n m_0^6(-dhw+dnl-cjn+cr_1w+ghj-glr_1)\,,\nonumber\\
	C_{v_2 a_3} &=&C_{v_3 a_2}=m m_0^6(-dhw+dnl-cjn+cr_1w+ghj-glr_1)\,,\nonumber\\
	C_{v_1 a_4} &=&n m_0^6(dlm-dr_2w+chw-cjm-ghl+gjr_2)\,,\nonumber\\
	C_{v_2 a_4} &=&m m_0^6(dlm-dr_2w+chw-cjm-ghl+gjr_2)\,,\nonumber\\
	C_{v_4 a_1} &=&
	n m_0^6(d(-w(h+m+r_2)+nl+lm)+cw(h+n+r_1)\nonumber\\
	&&-cj(n+m)+g(h(j-l)+jm+jr_2-nl-lr_1))\,, \nonumber\\
	C_{v_4 a_2} &=&
	m m_0^6(d(-w(h+m+r_2)+nl+lm)+cw(h+n+r_1)\nonumber\\
	&&-cj(n+m)+g(h(j-l)+jm+jr_2-nl-lr_1))\,,
\end{eqnarray}

\begin{eqnarray}
	\tilde{C}_{a_1 a_2} &=&-m_0^4(n+m)\,,\nonumber\\
	\tilde{C}_{a_1 a_3} &=&m_0^6(hn+n^2-r_1m)\,,\nonumber\\
	\tilde{C}_{a_1 a_4} &=&m_0^6(m(n+r_1)-hn)\,,\nonumber\\
	\tilde{C}_{a_2 a_3} &=&m_0^6(n(m+r_2)-hm)\,,\nonumber\\
	\tilde{C}_{a_2 a_4} &=&m_0^6(hm-r_2n+m^2)\,,\nonumber\\
	\tilde{C}_{a_3 a_4} &=&m_0^8(-2hnm+r_2n^2+m^2r_1)\,,
\end{eqnarray}

\begin{eqnarray}
	C_{a_1 a_1} &=&m_0^4(dnw+gjn-gr_1w)\,,\nonumber\\
	C_{a_2 a_2} &=&m_0^4(cmw+glm-gr_2w)\,,\nonumber\\
	C_{a_3 a_3} &=&(fr_1-2dj)(-2hnm+r_2n^2+m^2r_1)m_0^8/2\,,\nonumber\\
	C_{a_4 a_4} &=&(fr_2-2cl)(-2hnm+r_2n^2+m^2r_1)m_0^8/2\,,\nonumber\\
	C_{a_1 a_2} &=&m_0^4(dmw+cnw+g(-2hw+jm+nl))\,,\nonumber\\
	C_{a_1 a_3} &=&
	m_0^6(d(-hnw-2jnm+n^2l+mr_1w)+cjn^2+g(jmr_1-hjn))\,,\nonumber\\
	C_{a_1 a_4} &=&
	-(c(n(hw+jm-2nl)-mr_1w)+gl(hn-mr_1)\nonumber\\
	&&+dnlm+f(-2hnm+n^2r_2+m^2r_1))m_0^6\,,\nonumber\\
	C_{a_2 a_3} &=&
	m_0^6(f(-2hnm+n^2r_2+m^2r_1)+gj(hm-nr_2)\nonumber\\
	&&+d(hmw-2jm^2+nlm-nr_2w)+cjnm)\,,\nonumber\\
	C_{a_2 a_4} &=&m_0^6(-dlm^2+gl(hm-nr_2)+c(hmw-jm^2+2nlm-nr_2w))\,,\nonumber\\
	C_{a_3 a_4} &=&m_0^8(-dl-cj+fh)(-2hnm+r_2n^2+m^2r_1)\,.
\end{eqnarray}

\end{document}